		\documentclass[%
		 reprint,
	     showkeys,
		 amsmath,amssymb,
		 aps,
		]{revtex4-1}
		
		\usepackage{graphicx}
		\usepackage{dcolumn}
		\usepackage{bm}
		\usepackage{epstopdf}
		
	\usepackage{color}
\usepackage{mathtools}	
	
\begin{document}
	\definecolor{orange}{rgb}{1.0, 0.5, 0.0}
	\definecolor{purple}{rgb}{0.5, 0, 0.5}

			\let\vaccent=\v 
			\renewcommand{\v}[1]{\ensuremath{\mathbf{#1}}} 
			\newcommand{\gv}[1]{\ensuremath{\mbox{\boldmath$ #1 $}}}    
			\newcommand{\uv}[1]{\ensuremath{\mathbf{\hat{#1}}}} 
			\newcommand{\abs}[1]{\left| #1 \right|} 
			\newcommand{\avg}[1]{\left< #1 \right>} 
			\let\underdot=\d 
			\renewcommand{\d}[2]{\frac{d #1}{d #2}} 
			\newcommand{\dd}[2]{\frac{d^2 #1}{d #2^2}} 
			\newcommand{\pd}[2]{\frac{\partial #1}{\partial #2}} 
			\newcommand{\pdd}[2]{\frac{\partial^2 #1}{\partial #2^2}} 
			\newcommand{\pddd}[2]{\frac{\partial^3 #1}{\partial #2^3}} 
			\newcommand{\pdc}[3]{\left( \frac{\partial #1}{\partial #2}
			 \right)_{#3}} 
			\newcommand{\ket}[1]{\left| #1 \right>} 
			\newcommand{\bra}[1]{\left< #1 \right|} 
			\newcommand{\braket}[2]{\left<#1\vphantom{#2}\right| \left.#2\vphantom{#1}\right>} 
			\newcommand{\ketbra}[2]{\left|#1\vphantom{#2}\right> \left<#2\vphantom{#1}\right|} 
			\newcommand{\matrixel}[3]{\left< #1 \vphantom{#2#3} \right| #2 \left| #3 \vphantom{#1#2} \right>} 
			\newcommand{\commute}[2]{\left[#1,#2\right]} 
			\newcommand{\trace}[1]{\mathrm{Tr}\left[#1\right]} 
			\newcommand{\grad}[1]{\gv{\nabla} #1} 
			\let\divsymb=\div 
			\renewcommand{\div}[1]{\gv{\nabla} \cdot #1} 
			\newcommand{\curl}[1]{\gv{\nabla} \times #1} 
			\newcommand{\properint}[4]{\int_{#1}^{#2}#3\mathrm{d}#4}
			\newcommand{\improperint}[2]{\int_{-\infty}^{+\infty}#1\mathrm{d}#2}
		\preprint{APS/123-QED}
		
		\title{Quantum Energy Teleportation across a three-spin Ising chain in a Gibbs State}

		\author{Jose Trevison}%
		 \email{jose@tuhep.phys.tohoku.ac.jp}
		\author{Masahiro Hotta}
		 \email{hotta@tuhep.phys.tohoku.ac.jp}
		
		\affiliation{%
		 Graduate School of Science, Tohoku University \\Sendai 980-8578, Japan 
			}%
		
		
		
	
		
		\date{\today}
		
		\begin{abstract}
In general, it is important to identify what is the informational resource for quantum tasks.  Quantum energy teleportation (QET) is a quantum task, which attains energy transfer in an operational meaning by local operations and classical communication, and is expected to play a role in future development of nano scale smart grids. We consider QET protocols in a three-element Ising spin system with non periodic boundary conditions coupled to a thermal bath. The open chain is the minimal model of QET between two edge spins that allows the measurement and operation steps of the QET protocol to be optimized without restriction. It is possible to analyze how two-body correlations of the system, like mutual information, entanglement and quantum discord, can be resources of this QET at each temperature. In particular, we stress that quantum discord is not the QET resource in some cases even if arbitrary measurements and operations are available. 
		\end{abstract}
		
	
		\maketitle
		
		
		\section{\label{sec:level1} Introduction}
	
Quantum correlation has recently attracted much interest as resource of quantum tasks including quantum parameter estimation \cite{est1, est2, est3}, quantum teleportation \cite{telep} and quantum computing \cite{computing}.  There exist many kinds of quantum correlations. One of the most known is quantum entanglement.  Remote state transfer using quantum  teleportation requires nonzero amount of entanglement as resource. However it is often stressed that entanglement is not the only quantum correlation. The concept of quantum discord, which has been introduced on reference \cite{Discord} and \cite{Discord2}, can be nonzero even if entanglement exactly vanishes. Discord without entanglement is referred as quantum Dissonance. In general, it is important to identify the correlation as information resource for every quantum protocol.    

Quantum Energy Teleportation (QET) \cite{QET1, QET2, QET3, spin} is a quantum protocol, which attains energy transfer in an operational meaning. For quantum many-body systems, the interaction among subsystems generates quantum correlations in the ground states. Since energy density operators of the system do not commute with each other due to this interaction, uncertainty relations yield zero-point fluctuation of energy density in the ground states. With a shift of the energy such as the average energy density of the system is fixed as zero, the energy density fluctuates between the zero value. Thus we have negative energy density in quantum theory.  This zero-point energy density fluctuations of two separate subsystems A and B are quantum mechanically correlated. Hence, if we measure subsystem A and obtain the measurement result $\alpha$, this includes some information about energy density around subsystem B. During this measurement, positive amount of energy $E_A$ is injected into A, because the post-measurement state is not the ground state but an excited state. This property is called the  passivity of the ground state. By performing an appropriate local operation on B, dependent on  $\alpha$, the quantum fluctuation can be suppressed and negative energy density appears around B. Because of energy conservation,  positive amount of energy $E_B$ is extracted by the operation device on B. Note that B is in a state with zero-energy before the measurement, thus extraction of $E_B$ looks like energy extraction from nothing. This is QET. $E_A$ is regarded as input energy and $E_B$ as output energy of QET. Non-negativity of total Hamiltonian imposes $E_A\geq E_B$ \cite{Hotta}. 

The QET protocol has attracted attention for future development of nano-scale smart grids with low power consumption \cite{Smart}. An experiment to verify QET using edge channel currents in a quantum Hall system has been proposed \cite{Hall}.  QET is also related with information thermodynamics.  Quantum Maxwell's demons, by using the measurement results, are able to adopt QET in order to extract more energy out of quantum systems \cite{Hotta}.  In black hole physics, QET plays a crucial role. By measuring zero-point fluctuation of quantum fields outside horizon and performing QET, the area of event horizon shrinks and its black hole entropy decreases \cite{BH}. Even in low temperature cases, QET remains effective. In Gibbs states below a critical temperature, many-body systems which have ground states with maximum-rank entanglement structure, satisfy strong local passivity \cite{strong}. This means that positive amount of energy is injected during arbitrary local operations. Therefore, we are not able to extract thermal energy from subsystem B only by local operations. In the low-temperature regime, thermal energy extraction requires global operations like time evolution generated by Hamiltonian. However, if we adopt a QET protocol, the passivity is broken and a part of thermal energy can be extracted \cite{Frey}. 

As conventional quantum teleportation, it is of importance to understand what correlation is the resource of QET.  For the case of the ground state, it is known that quantum entanglement is the resource of QET \cite{Hotta}. However, the finite-temperature case is nontrivial. In a simple toy model with a two qubit chain with non-demolition measurements of the interaction Hamiltonian, quantum discord is seen to act as a resource for QET at high temperatures \cite{Frey}. However, so far other quantum systems have not been explored.  In addition, the two-spin model imposes a severe limitation on QET optimization problem of local measurements of energy-sender qubit.  Only non-demolition measurements are available, which do not disturb the potential term between the two spins. Thus there exists no consensus about the resource in general cases beyond the two qubit system. 

To avoid the limitation of available measurements, we consider a three-spin open chain model and QET from one edge spin to another edge spin in this paper. The optimal measurements of energy-sender spin and optimal local unitary operations of energy-receiver spin, which provide the maximal amount of teleported energy, are determined in the ground-state case and finite-temperature case. The maximum teleported energy is compared to various two-body correlations between two edge spins, including quantum mutual information, quantum discord \cite{Discord, Discord2}, concurrence \cite{Concurrence}, and negativity \cite{Negativity}. In contrast with the two qubit chain, we show that quantum discord is not a perfect resource of QET in the three qubit system. Through the variations of one of the parameters in the model, we found that the optimal teleported energy becomes zero in some cases at zero and finite temperature, where entanglement vanishes, but quantum discord is still present. This implies that quantum discord cannot become QET resource of this system at this regime. In other words, we found a regime where there is quantum discord  between the edge spins but energy teleportation is not possible. This is an unexpected result and crucial for the identification problem of quantum task resource.

Because through the variation of one parameter in the model it is possible to modify the amount of teleported energy, such as it is impossible to have QET; we pose an analogy between energy transportation in the three spin
chain model and field effect transistors (FET).  This paper is structured as follows. Section II introduces our three-particle model and also outlines the derivation of the teleported energy when using projective measurements; the supporting calculations are given in the appendices. Section III explores the relationship between the QET protocol's efficiency and the degree and type of quantum correlations in the system.  Section IV contains a detailed study, using the most general measurements that can be applied to the system, of the regime in which no teleported energy is possible, even though there are quantum correlations. Section V offers our conclusions.

			\section{\label{sec:level1} Three Qubit Model}
	
	Consider a system of three spin $1/2$ particles, whose labels are 1, 2 and 3 respectively, with Hamiltonian: 
			\begin{equation}
			H=\kappa  \left(\sigma _{1,x}\sigma_{2,x}+\sigma_{2,x}\sigma_{3,x}\right)+\sigma_{1,z}+\lambda \ \sigma _{2,z}+\sigma_{3,z}
			\label{Hamiltonian}
			\end{equation}
		
			The model resembles  the Ising chain model in a transverse magnetic field for only three elements in the chain,  without the interaction term corresponding to the periodic boundary conditions associated to a ring topology. The $\sigma$ operators are the Pauli operators, $\kappa$ is a dimensionless parameter that takes real values and represents the relative strength of the coupling between spins and the intensity of the magnetic field, and $\lambda$, also a dimensionless parameter, represents the strength of the coupling of the spin of particle 2 with the transverse magnetic field.  The presence of these two parameters in the model allow us to study the two main factors in order to teleport energy; the amount of correlations between subcomponents of the total system, and the capability to generate local negative energy density around subsystem B after Bob's operation. If we restrict Bob to perform measurements on particle 3, then it's local Hamiltonian  is independent of $\lambda$
\begin{equation}
H_B=\kappa \ \sigma_{2,x}\sigma _{3,x} +\sigma_{3,z}
\label{BobLocal}
\end{equation}

Therefore, though the parameter $\lambda$ is unrelated to the generation of local negative energy at particle 3, it is related to the correlation between particles 1 and 3. In fact, for $\lambda \rightarrow \infty$ the total Hamiltonian reduces to the $z$ component of particle 2, meaning, in particular, that particles 1 and 3 become uncorrelated. The parameter $\lambda$ also dictates the degeneracy of the system; when $\lambda = 0$ the system is completely degenerated (because in this case the Hamiltonian (\ref{Hamiltonian}) commutes with $\sigma_{2,x}$) and non-degenerate otherwise. On the other hand, the $\kappa$ parameter is related to the generation of local negative energy, since it is present on Bob local Hamiltonian (\ref{BobLocal}) and to the correlations since physically represents the coupling between spins, in other words, in the limit $\kappa=0$ there are not correlations.  Specifically, fixing a value of $\kappa$ while changing the parameter $\lambda$ is equivalent to study the effect of only the correlations of the system in the energy teleportation.     

 Since the Hamiltonian is dimensionless, so are the eigenergies [Appendix \ref{Appendix}].  If the three-particle system described by equation (\ref{Hamiltonian}) is weakly coupled to a thermal bath with temperature \textit{T}, the state of the system will be given by a Gibbs state [Appendix \ref{Appendix}]. Since the Eigenergies are dimensionless, so it is the temperature parameter \textit{T}.  

Let the three-particle system be partitioned into two subsystems, A and B, such that Alice and Bob, respectively, can make measurements on (just) A and B.  The QET protocol can be applied to the system as follows:	

	Alice performs a projective measurement on subsystem A with output $\alpha$ ($\pm1$). 
\begin{eqnarray}
	&&		M_A(\alpha)=\frac{1}{2} \left(\mathcal{I}_A+\alpha \ \uv{r}_A \cdot \gv{\sigma}_A\right)
\label{projective}
\\ &&
  \uv{r}_A=\left(\sin{(\theta)}\cos{(\phi)},\ \sin{(\theta)}\sin{(\phi)}, \ \cos{(\theta)}\right) 
		\end{eqnarray}

	Where the label A can refer to qubit 1, 2 or 3 and $\uv{r}_A$ is a unit vector to be chosen such as the maximum amount of teleported energy is achievable. Due to the strong passivity of the ground state, this measurement injects energy $E_A$ into the system. On the other hand, for a finite temperature \textit{T}, the Gibbs states of a finite quantum system are strong locally passive, in other words; any local operation, measurement included, will introduce further energy into the system below a critical temperature specific to the system and subsystem \cite{strong}. An additional condition for Alice operations is necessary in order to avoid direct energy input into Bob's system. 
  	\begin{eqnarray}
	&& \commute{M_A(\alpha)}{V_{AB}}=0
    \label{restriction}
\end{eqnarray}

		 Where $V_{AB}$ is the interaction term between Alice and Bob system. Note that in the case Alice measures only qubit 1, and Bob qubit 3, the equation (\ref{restriction}) is satisfied by the most general measurement $M_A$, since there is no interaction term $V_{13}$ on the Hamiltonian (\ref{Hamiltonian}). This is the main difference between the three qubit and the two qubit system, and is the case in which the results of Section III will be based on.  On section IV we will work with more general measurements involving the two qubits 1 and 2.   

On the the second step of the QET protocol, Alice announces the measurement result $\alpha$ to Bob via a classical channel. The time evolution of the system was neglected since it was assumed that the speed of the communication is greater than the energy diffusion of the system.
		
On the final step of the QET protocol; Bob performs at subsystem B a local operation $U_B(\alpha)$ dependent on Alice's result $\alpha$. The operation $U_B(\alpha)$ is chosen such as it generates negative energy $-E_B$ on subsystem B, which is equivalent by energy conservation that the measurement yields a positive energy $+E_B$. With the eventual cool down of the system the input energy $E_A$ will compensate the local negative energy $-E_B$.
		 \begin{eqnarray}
	&& U_B(\alpha)=\exp{\left[-\imath\alpha (\v{r}_B \cdot \gv{\sigma_3})\right]}
			\\&&
			\v{r}_B= r\left(\sin{(\delta)}\cos{(\gamma)},\ \sin{(\delta)}\sin{(\gamma)}, \ \cos{(\delta)}\right) 
		\end{eqnarray}

Where $\v{r}_B$ is a vector chosen such as the maximum amount of teleported energy is achievable. In general, the label B can refer to qubit 1, 2 or 3, with naturally $A\neq B$; however from now on Bob will be restricted to perform measurements only on qubit 3.   
	
	If the system is in equilibrium with a thermal bath at temperature \textit{T}, the state of the system will be given by a Gibbs state $\rho=\rho(\textit{T})$ [Appendix \ref{Appendix}]; then the average energy input $E_A$ by Alice's measurement is given by: 
\begin{equation}
E_A=\sum_\alpha\trace{M_A^\dagger(\alpha) H M_A(\alpha)\rho} - \trace{H \rho}
\end{equation}

On the other hand; the average energy loss of the system after the QET protocol, in other words the average amount of teleported energy $E_B$, can be calculated as the difference between  the average energy after Alice's operation and after Bob's operation. 
\begin{equation}
\begin{multlined}
		E_B= \sum_\alpha\trace{M_A^\dagger(\alpha) H M_A(\alpha)\rho} 
	\\ \ \ \  \ \ 
	-\sum_\alpha\trace{U_B^\dagger(\alpha) M_A^\dagger(\alpha) H M_A(\alpha) U_B(\alpha) \rho}
\\ \
=\sum_\alpha \trace{M_A^\dagger(\alpha)M_A(\alpha)U_B^\dagger(\alpha)\commute{U_B(\alpha)}{H_B}\rho} 
\label{QET-formula}
\end{multlined}
\end{equation}

The quantity $E_B$ defined in the previous equation can be positive or negative; only in the case it is positive we will have teleported energy. For the moment, let us assume that the unitary operation $U_B(\alpha)$ is such as $E_B$ is a positive quantity; then in order to qualify the effectiveness of the QET protocol, we introduced the efficiency $\eta$, which  allows us to compare dimensionless amount of energies, without specifying the energy scale of the system.  
\begin{equation}
\eta=100 \times \frac{E_B}{E_A}
\end{equation}		

\section{Projective Measurements Results}

The main result of this paper was obtained for the case in which we restrict Alice's measurement to qubit 1, and Bob's measurement to qubit 3. By using the results  of the Appendix [\ref{Appendix}], for the functions $\mathcal{A}, \mathcal{R}, \mathcal{B}$, functions of the Temperature \textit{T} and the parameters $\kappa, \lambda$; the teleported energy can be written as: 
\begin{eqnarray}
&&
E_B=A \sin{(2r)}-B (1-\cos{(2r)})
\label{QEToptimal}
\\ 
 &&
A= \mathcal{A} \sin(\theta) \sin(\delta) \sin(\phi-\gamma)
\\ &&
B=\sin^2{(\delta)}\left[2\mathcal{B} \cos^2{(\gamma)}+\mathcal{R}\right]-2\mathcal{B}
\end{eqnarray}

The optimization of $E_B$ in (\ref{QEToptimal}), such as the maximum amount of teleported energy is obtained, was done by the criteria of the second partial derivatives and the Hessian matrix, together with numerical calculations. The maximum amount of teleported energy is obtained when $\theta=\pi/2, \phi=\pi/2$ which implies a projective measurement of $\sigma_{1,y}$ by Alice, and $\delta=\pi/2, \gamma=0$ which implies  that the unitary operation of Bob is related to $\sigma_{3,x}$. Substituting those values on equation (\ref{QEToptimal}), and optimizing the last parameter r:
\begin{equation}
\tan{\left(2 r_0\right)}=\frac{\mathcal{A}}{\mathcal{R}}
\end{equation}

Then the maximum amount of teleported energy $E_B^{max}$ is given by:
\begin{equation}
E_B^{max}=\sqrt{\mathcal{A}^2(\lambda,\kappa,T)+\mathcal{R}^2(\lambda,\kappa,T)}-\mathcal{R}(\lambda,\kappa,T)
\label{MAX}
\end{equation}

Since a separable ground state leads to zero teleported energy \cite{Hotta}, it is thought that quantum correlations, particularly entanglement, are necessary in order to have teleported energy. Therefore the quantum correlations must be studied in detail. Since Alice and Bob are restricted to the edge spins 1 and 3 respectively, then let us focus on the quantum correlations between the particles 1 and 3. Considering only the edge spins the reduced state of the system can be written as $\rho_{13}=\mathrm{Tr}_2\left[\rho\right]$, with this density operator the concurrence \cite{Concurrence} and the negativity \cite{Negativity} were calculated. The detail calculation of the quantum correlations can be found on the appendix [\ref{Appendix3}]. The concurrence $C(\rho_{13})$, is a measure of quantum entanglement \cite{Concurrence}, that can be calculated in terms of the eigenvalues $\Lambda_{j,C}$ of $\rho_{13} {\tilde \rho_{13}}$, where $\tilde\rho_{13}=\left(\sigma_1^\mathcal{Y}\otimes\sigma_3^ \mathcal{Y}\right)\rho_{13}^\star\left(\sigma_1^\mathcal{Y}\otimes \sigma_3^\mathcal{Y}\right)$. 
\begin{equation}
C(\rho_{13})= \text{max} \left\{\sqrt{\Lambda_{2,C}}-\sqrt{\Lambda_{1,C}}-\sqrt{\Lambda_{3,C}}- \sqrt{\Lambda_{4,C}}, 0 \right\}
\end{equation}

Another simple to calculate entanglement measure is the Negativity $\mathcal{N}(\rho_{13})$ \cite{Negativity}. This one is defined as the absolute sum of the negative eigenvalues  $\Lambda_{i,N}$ of $\rho_{13}^{(T_1)}$ the partial transpose of the density matrix $\rho_{13}$ with respect to qubit 1. 
\begin{equation}
\mathcal{N}(\rho_{13})=\sum_i \frac{\abs{\Lambda_{i,N}}-\Lambda_{i,N}}{2}
\end{equation}

However since both measurements only quantify the amount of entanglement, and there are other quantum correlations different from entanglement, we also calculated the discord \cite{Discord, Discord2}. Quantum discord measures the totality of the quantum correlation, entanglement and otherwise, between two components of a quantum system.  The range of the discord as a tool to study quantum systems cover physics like: the Heisenberg uncertainty principle \cite{Heisenberg}, quantum key distribution protocols \cite{QKD}, Heisenberg chains in a magnetic field \cite{chain}, etc. In the case there is not entanglement, the discord is called quantum dissonance. It has been proved that for some special cases of an assisted optimal state discrimination  only dissonance is required \cite{dissonance}.   

The discord associated with measurement on particle 1 can be calculated analytically [appendix \ref{Appendix3}]. In mathematical terms the quantum discord ${D}_{13}(\rho_{13})$ is defined as the difference of the quantum mutual information $I(\rho_{13})$, which contains all the classical and quantum correlations between the subsystems, and the classical correlations $J(\rho_{13})$, which contains all the information that can be obtained through local measurements on one of the subsystems. If $S(\rho)=-\trace{\rho \log\left(\rho\right)}$ is the Von Neumann Entropy, then the Quantum Discord $D_{13}(\rho_{13})$
\begin{eqnarray}
&&{D}_{13}(\rho_{13})=I(\rho_{13})-\text{max}_{\hat{\Pi}_1}\left\{J(\rho_{13})\right\}
\\ &&
I(\rho_{13})=S(\rho_{1})+S(\rho_{3})-S(\rho_{13})
\\ &&
J(\rho_{13})=S(\rho_1)-S(\rho_{13 \vert \hat{\Pi}_1^\alpha})
\end{eqnarray}

\begin{figure}[h]
 \includegraphics[width=4.2cm, height=3.3cm]{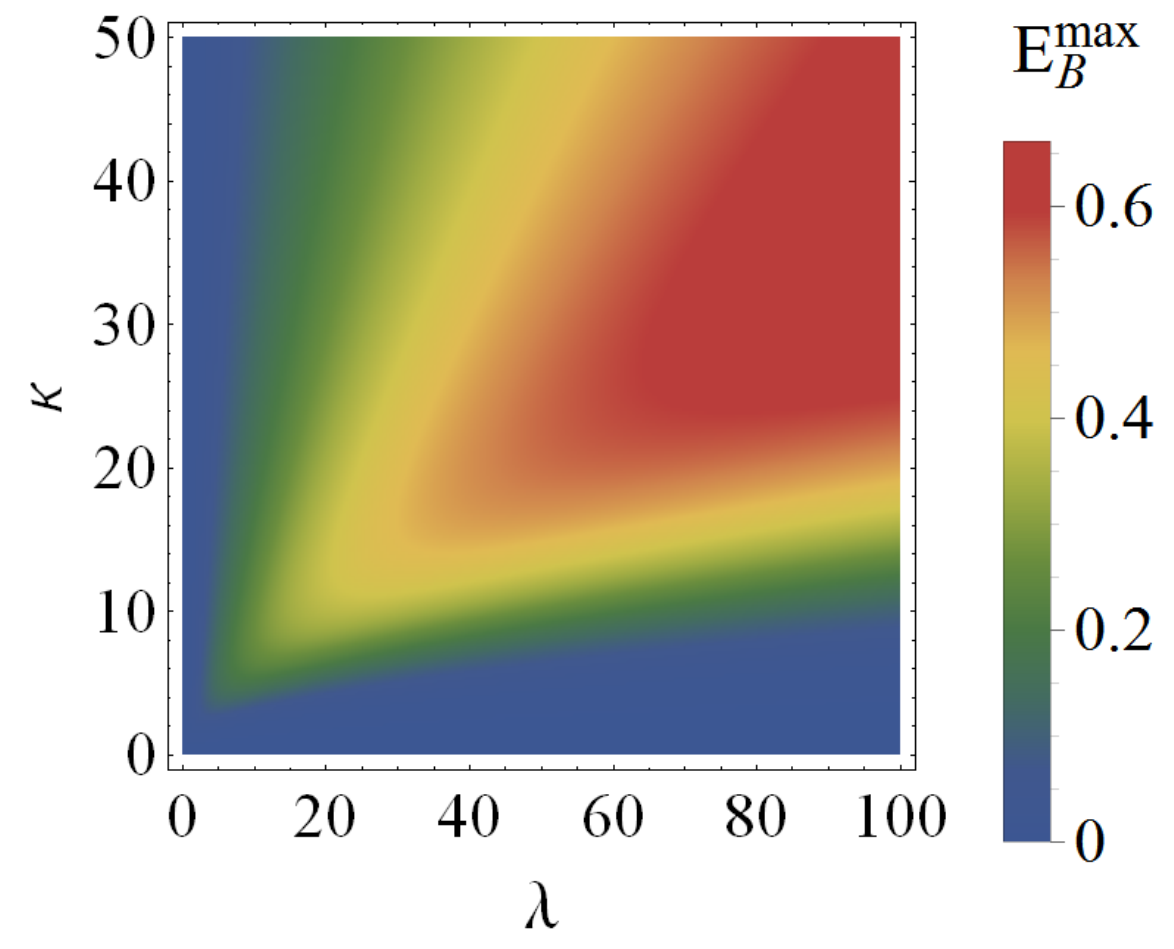}
\includegraphics[width=4.275cm]{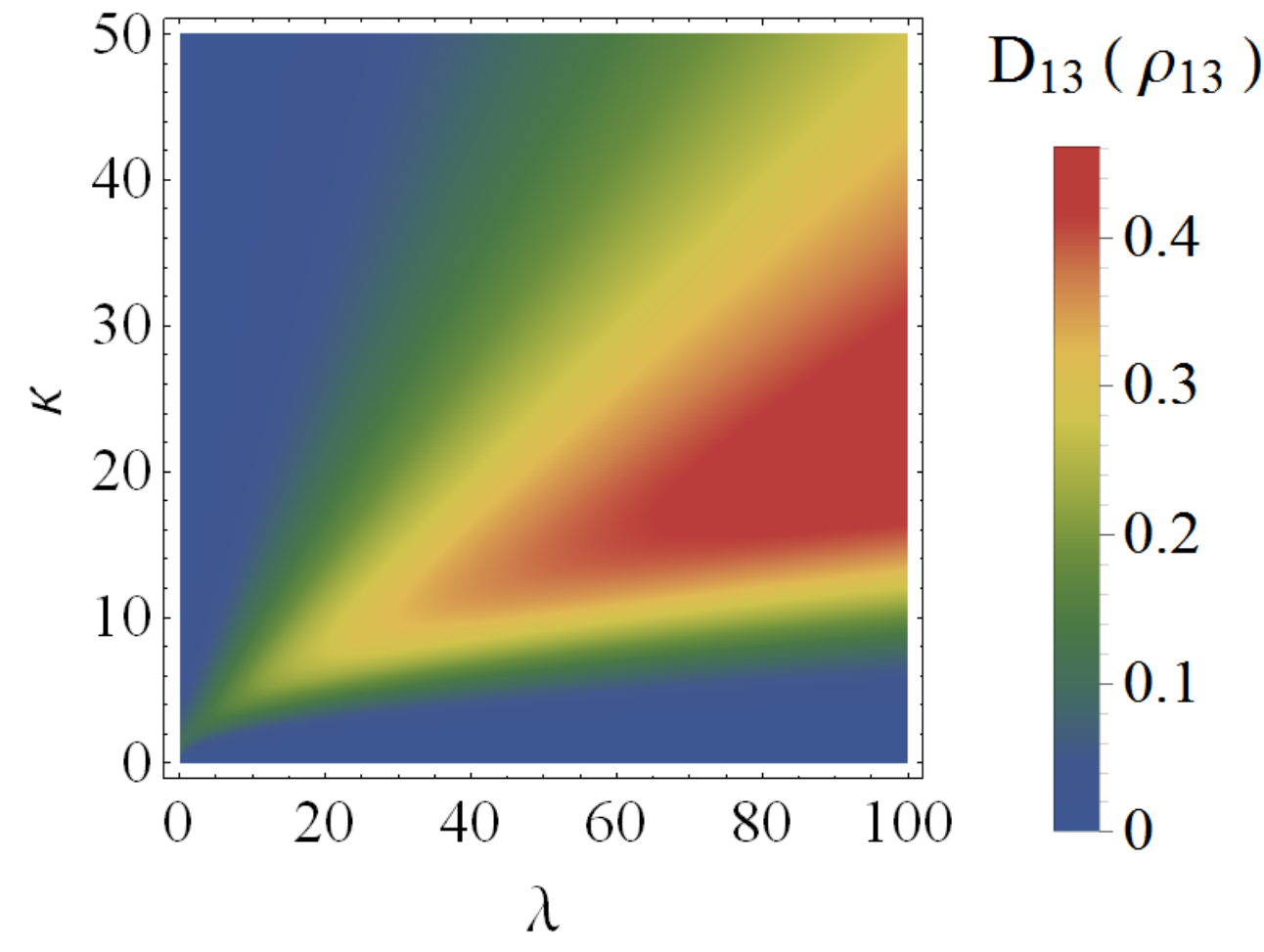}
\includegraphics[width=4.275cm]{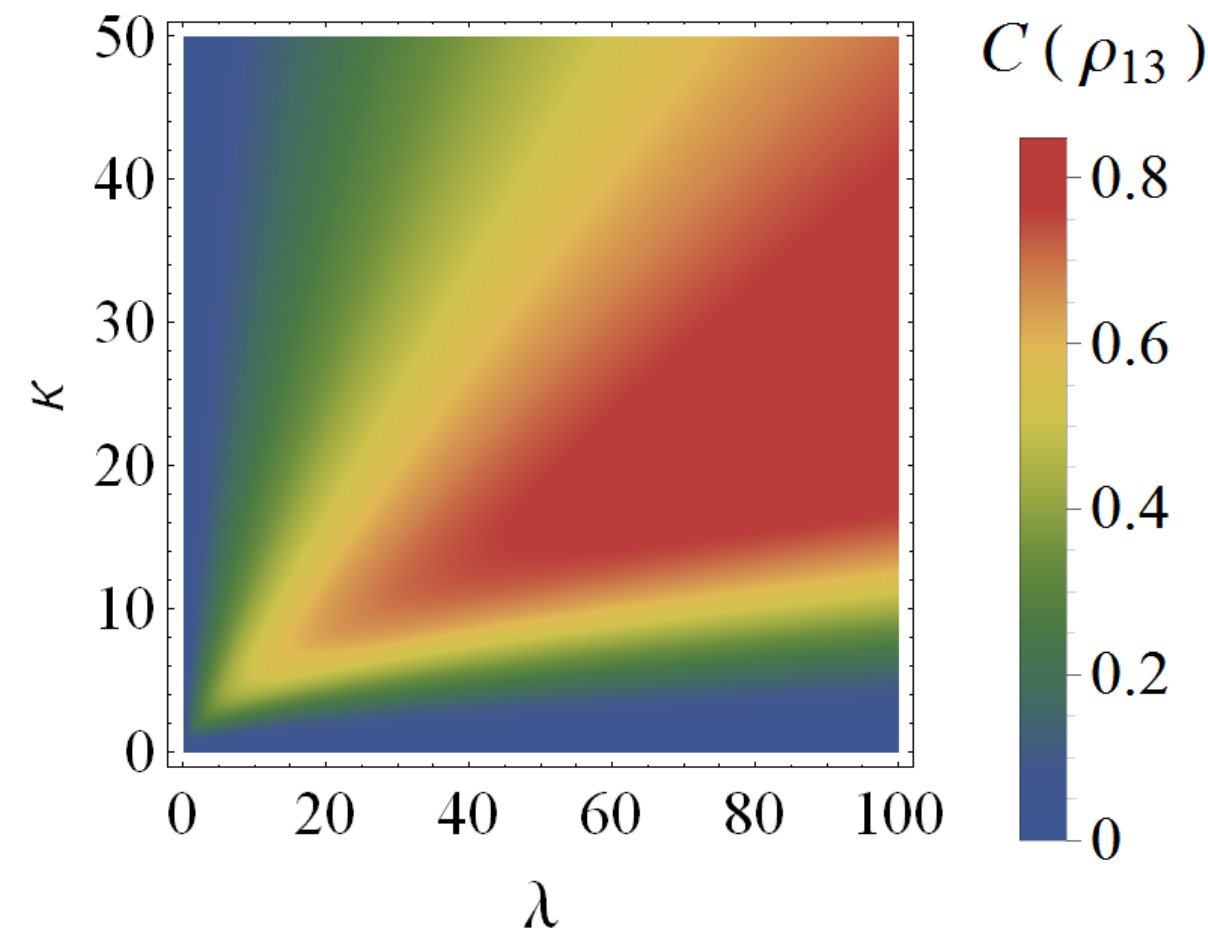}
\includegraphics[width=4.275cm]{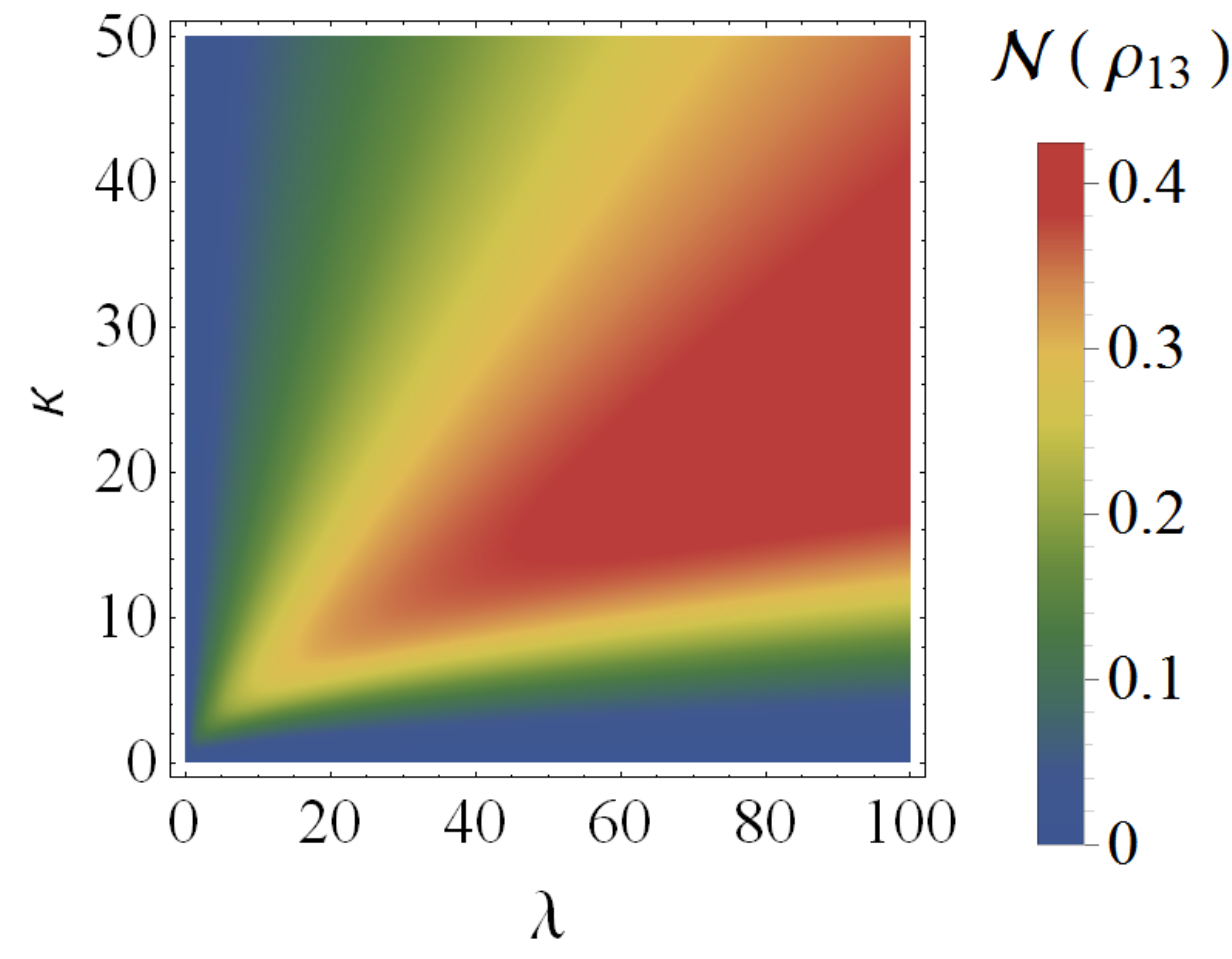}
	\caption{For \textit{T}=0; teleported energy $E_B^{max}$  (top left), quantum discord ${D}_{13}(\rho_{13})$ (top right), concurrence $C(\rho_{13})$ (bottom left) and negativity $\mathcal{N}(\rho_{13})$ (bottom right)}
\label{compare}
\end{figure}

Figure \ref{compare} compares the maximum teleported energy and the quantum correlations in the ground state (\textit{T}=0). The expressions for the quantum correlations can be seen in appendix [\ref{Appendix3}]. The teleported energy $E_B^{max}$, the quantum discord ${D}_{13}(\rho_{13})$, the concurrence $C(\rho_{13})$  and the negativity  $\mathcal{N}(\rho_{13})$ show the same behavior. In the no-coupling limit $\kappa\rightarrow0$, in which no quantum correlation exists, no energy teleportation is possible. 

On the other hand, when  $\lambda=0$,  the Hamiltonian of the system (\ref{Hamiltonian}) commutes with the operator $\sigma_{2,x}$, therefore all the eigenvalues are degenerated [appendix \ref{Appendix}]. Taking explicitly this limit on equation (\ref{MAX}),  since the function $\mathcal{R}$ is always positive for every value of $\lambda, \kappa$ and \textit{T}, and $\mathcal{A}(0,\kappa,\textit{T})=0$ [appendix \ref{Appendix}]; then when $\lambda=0$ it is not possible to teleport energy for the case of the ground state and also for a finite temperature case. In particular for the ground state, a detailed study of the negativity $\mathcal{N}(\rho_{13})$ and the concurrence $C(\rho_{13})$ shows that for $\lambda=0$ there is not entanglement between qubits 1 and 3 [Appendix \ref{Appendix3}]. In other words, even though there is dissonance, quantum correlations without entanglement, and we perform the most general projective measurements that can be applied to the system; it is not possible to teleport energy in the limit $\lambda=0$, since the ground state is not entangled \cite{Hotta}. 

\begin{figure}[h]
	\includegraphics[width=4.275cm]{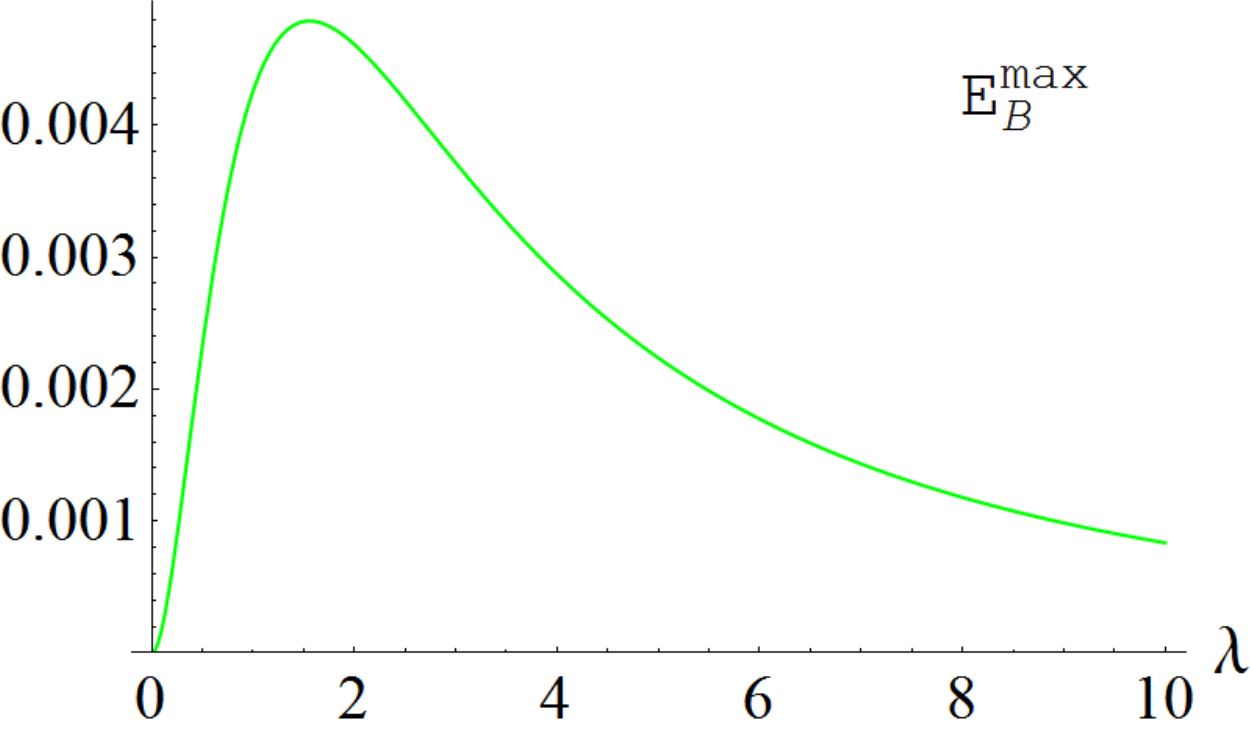}	
\includegraphics[width=4.275cm]{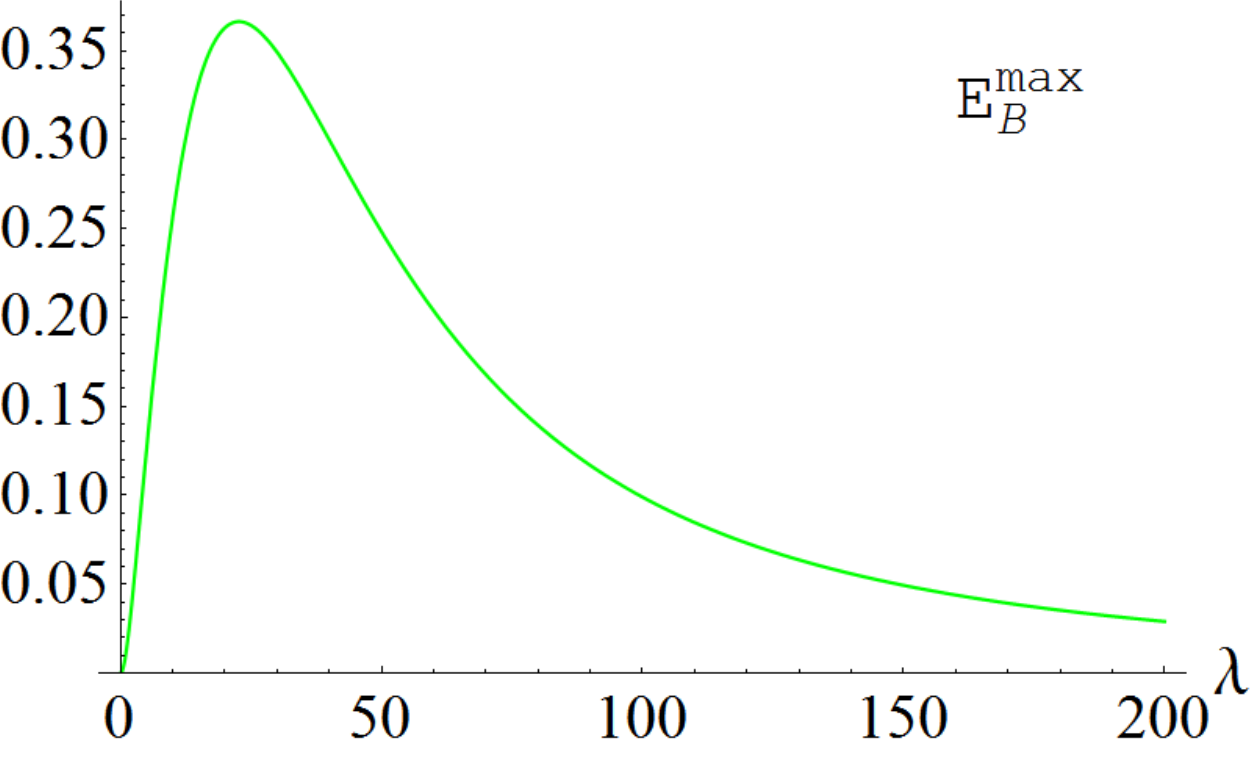}	
     \includegraphics[width=4.275cm]{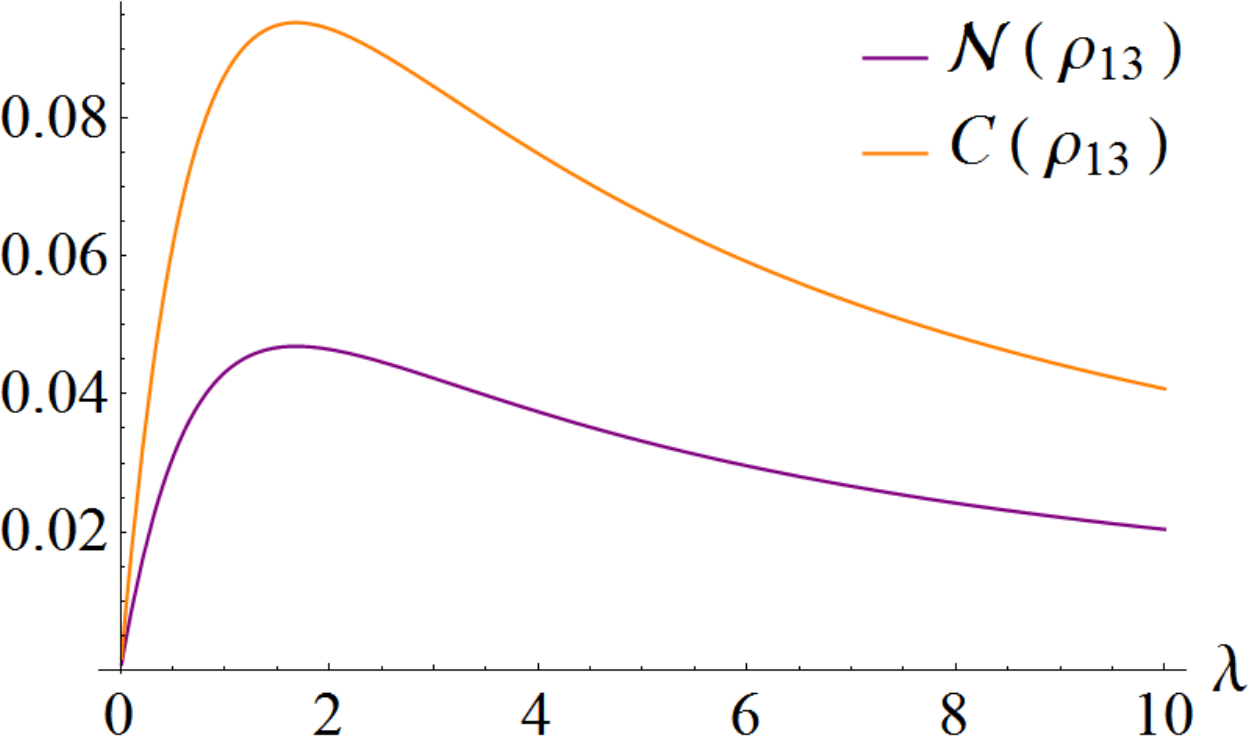}   
\includegraphics[width=4.275cm]{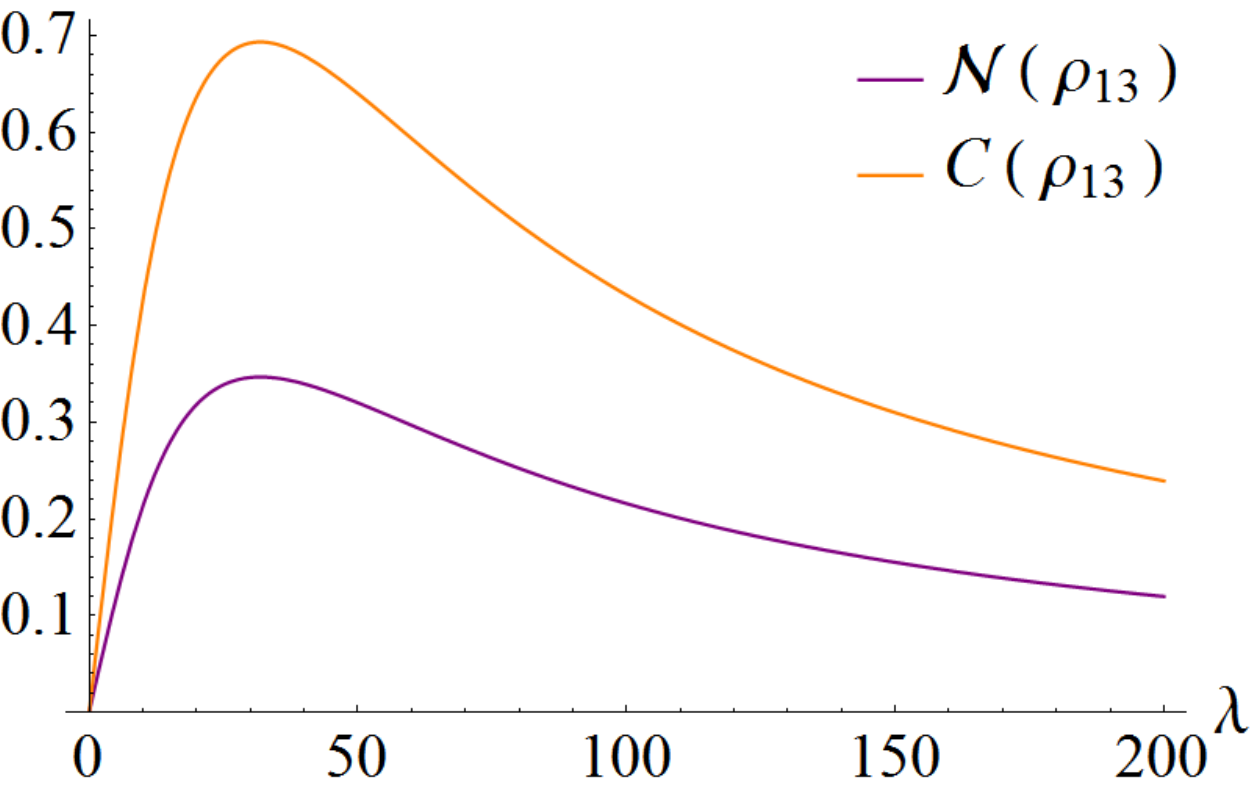}   
\includegraphics[width=4.275cm]{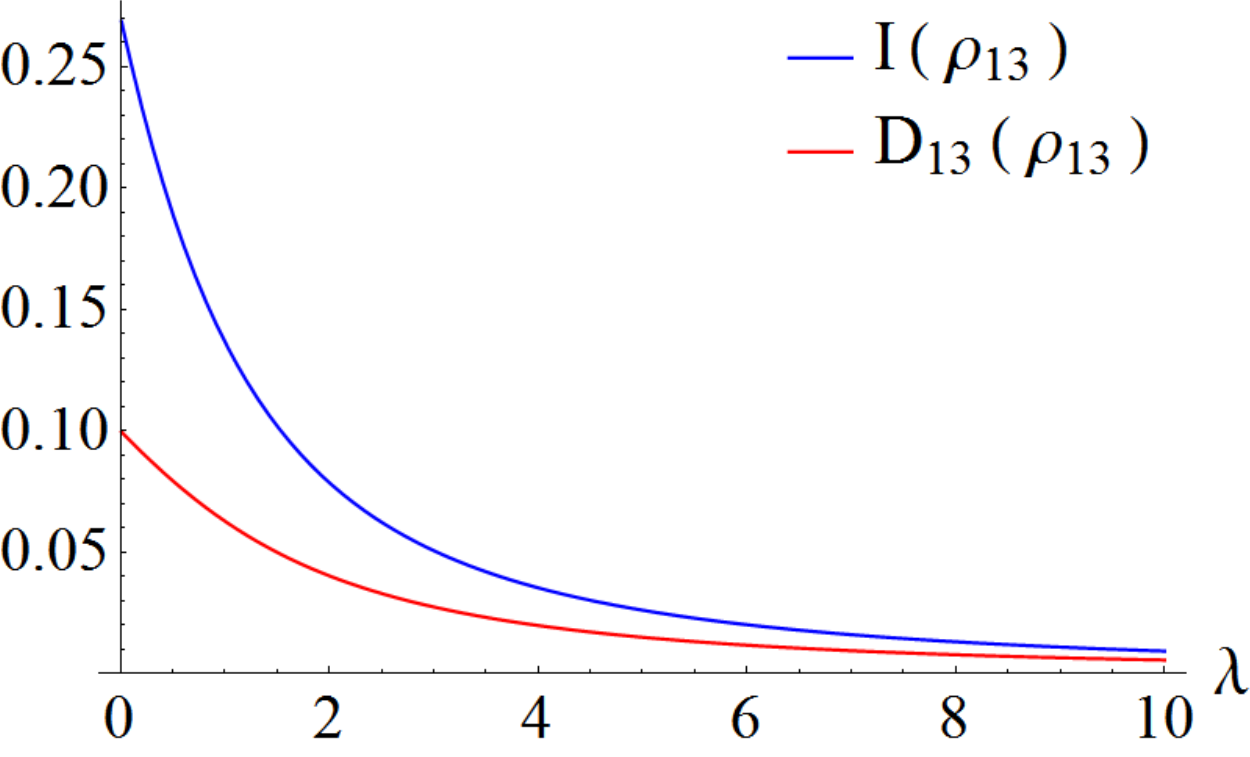}
\includegraphics[width=4.275cm]{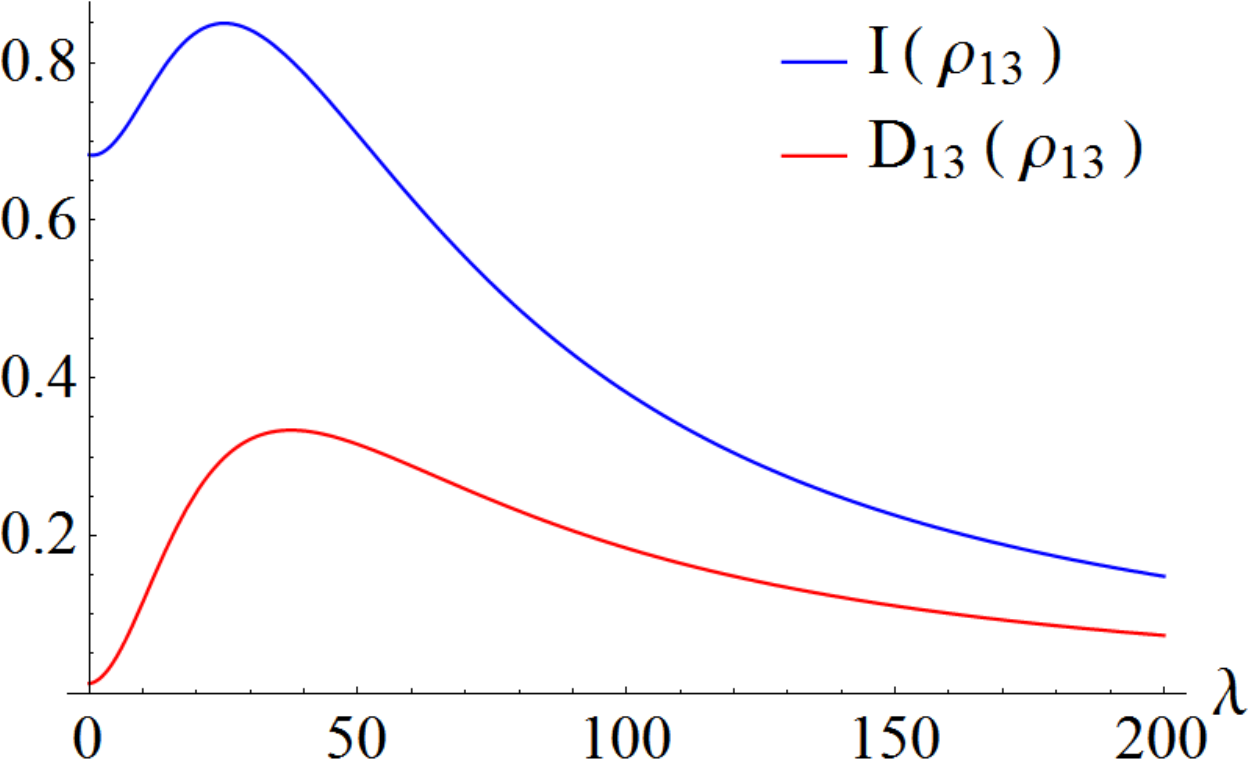}
	\caption{To the left: for \textit{T}=0 and $\kappa=1$ teleported energy $E_B^{max}$ (top), measurements of entanglement (middle) negativity $\mathcal{N}(\rho_{13})$ (purple) and concurrence  $C(\rho_{13})$ (orange); (bottom) mutual information I($\rho_{13}$) (blue) and quantum discord $D_{13}(\rho_{13})$ (red); to the right: for T=0 and $\kappa=10$ same quantities as in the left column}
\label{new2}
\end{figure}

On figure \ref{new2} the behavior of the negativity $\mathcal{N}(\rho_{13})$ and the concurrence $C(\rho_{13})$, as a function of the $\lambda$ parameter for \textit{T}=0 and two different values of the coupling parameter $\kappa=\{1,10\}$, among the teleported energy $E_B^{max}$ and the quantum discord ${D}_{13}(\rho_{13})$ and mutual information I($\rho_{13}$), can be found. In contrast with the negativity, concurrence, and the teleported energy, the discord and the mutual  information are different from zero when $\lambda=0$. Also it can be seen that a large amount of teleported energy is associated to larger values of the coupling parameter $\kappa$, limited to a non large $\lambda$, in other words a region of the phase space in which the interaction term of the Hamiltonian (\ref{Hamiltonian}) and the $\lambda$ term have meaningful contributions that are not overlapped between each other. 

\begin{figure}[h]
 \includegraphics[width=4.2cm, height=3.3cm]{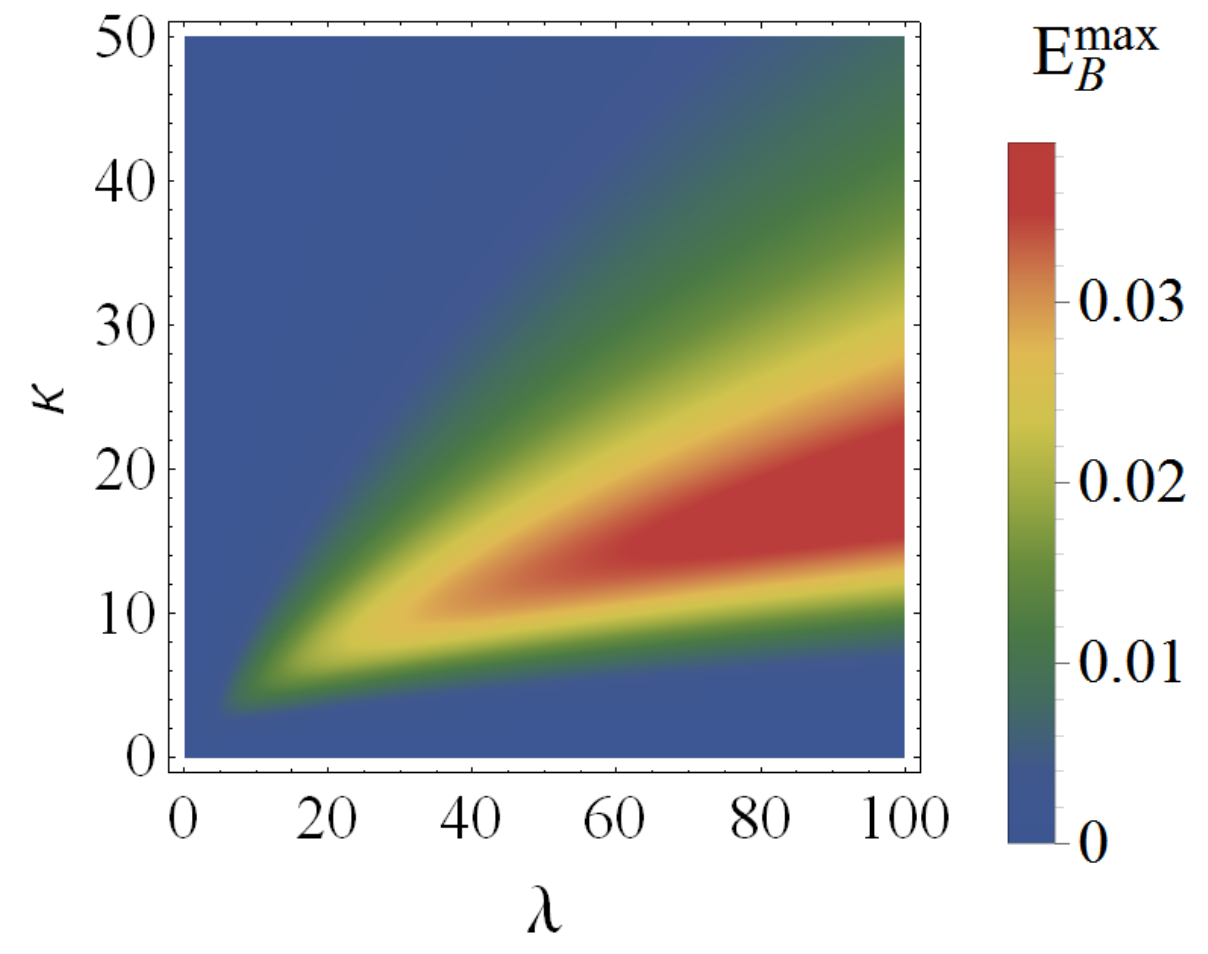}
 \includegraphics[width=4.275cm]{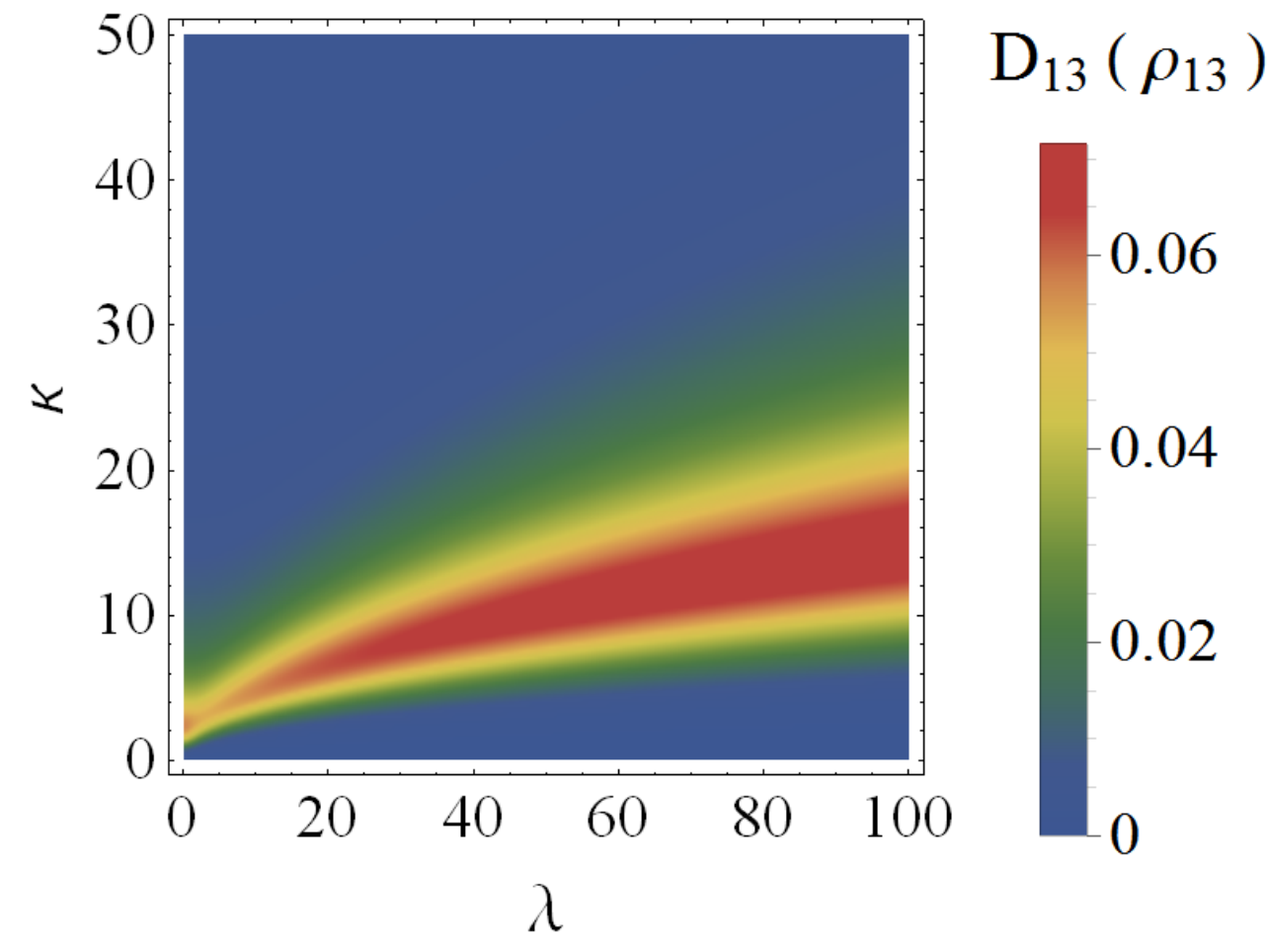}
\includegraphics[width=4.275cm]{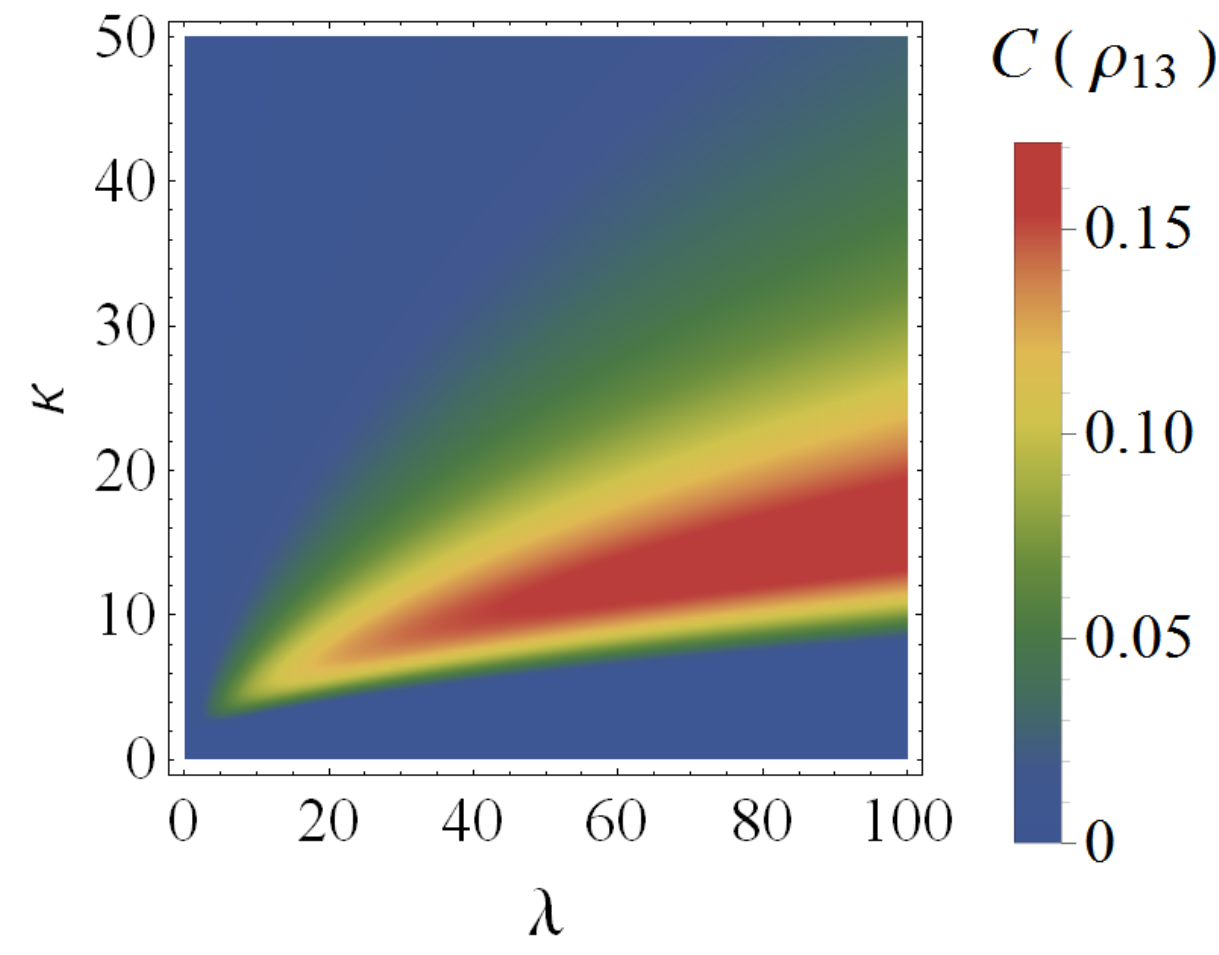}
 \includegraphics[width=4.275cm]{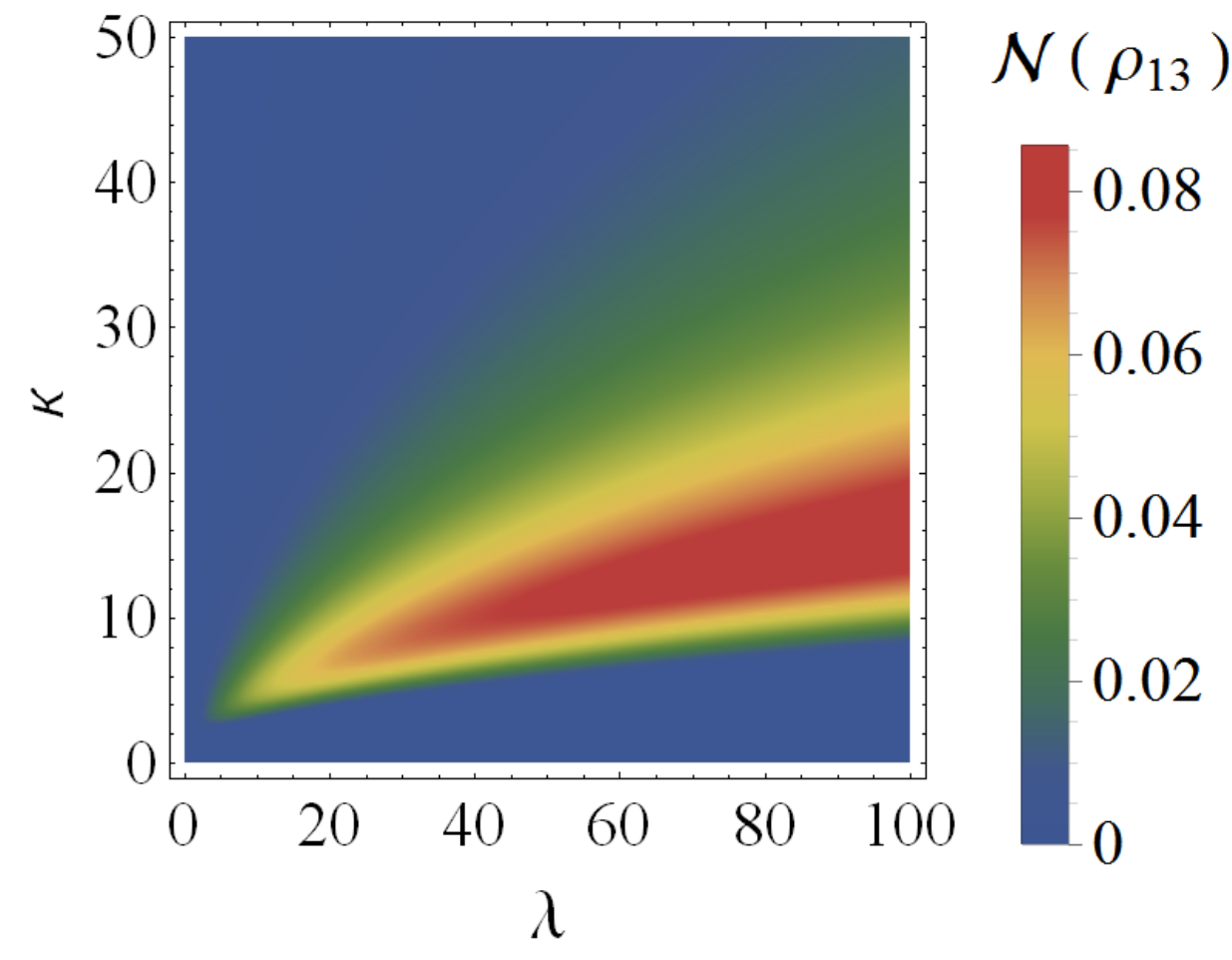}
	\caption{For \textit{T}=1; teleported energy $E_B^{max}$  (top left), quantum discord ${D}_{13}(\rho_{13})$ (top right), concurrence $C(\rho_{13})$ (bottom left) and negativity $\mathcal{N}(\rho_{13})$ (bottom right)}
\label{compare3}
\end{figure}

For the case of finite temperature, figure \ref{compare3} shows: teleported energy, quantum discord, concurrence and negativity for Temperature \textit{T}=1.  Similar to the case of the ground state, it is not possible to teleport energy in the limit $\kappa=0$, no correlations, and in the limit $\lambda=0$. Furthermore, similar to the case of \textit{T}=0; in the region of large $\lambda$, and small $\kappa$, in other words the region in which the Hamiltonian of the system (\ref{Hamiltonian}) can be approximated to be only the z component of particle 2, the value of teleported energy is reduced due to the decrease of the entanglement of the subsystem.  As can be seen by comparing with the case of \textit{T}=0 (figure \ref{compare}), the phase space, in which it is possible to obtain the larger values of teleported energy, shrinks with the increase of the temperature. This is a consequence of the decrease of the correlations, specifically entanglement,  a natural behavior for quantum correlations. 

\begin{figure}[h]
	\includegraphics[width=4.275cm]{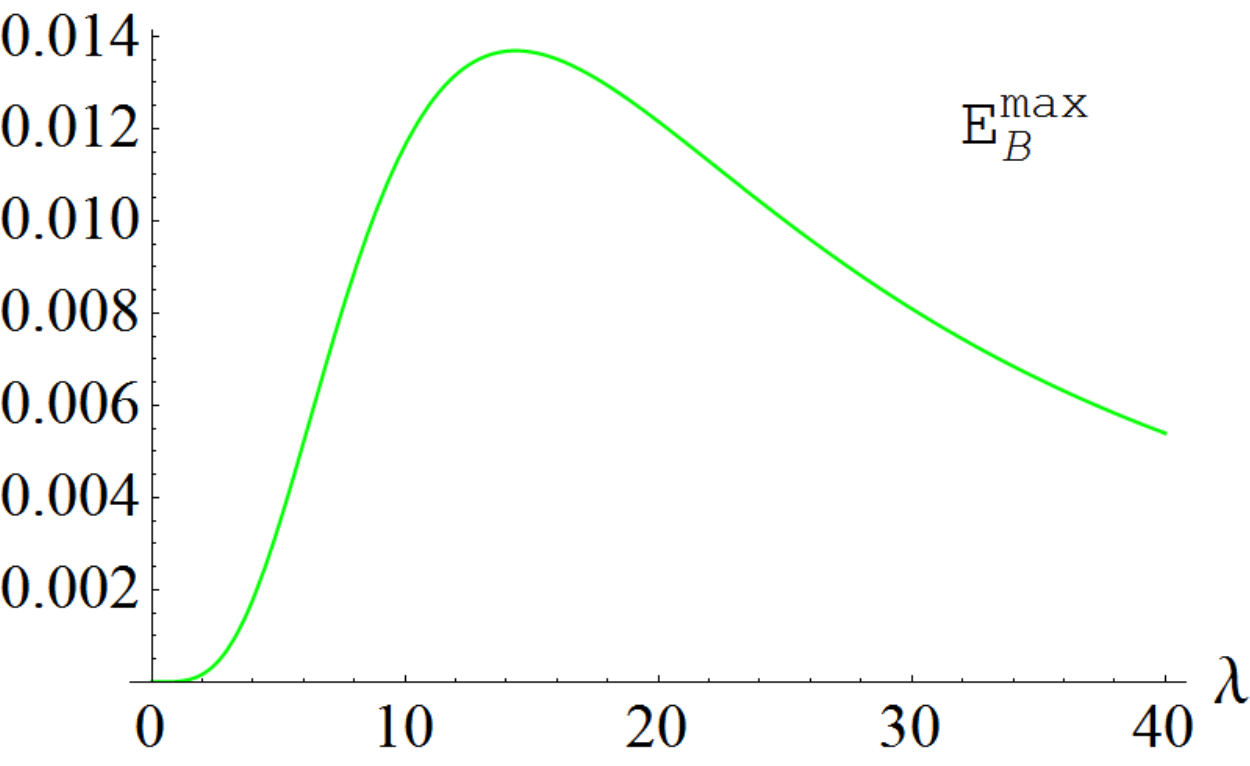}
	\includegraphics[width=4.275cm]{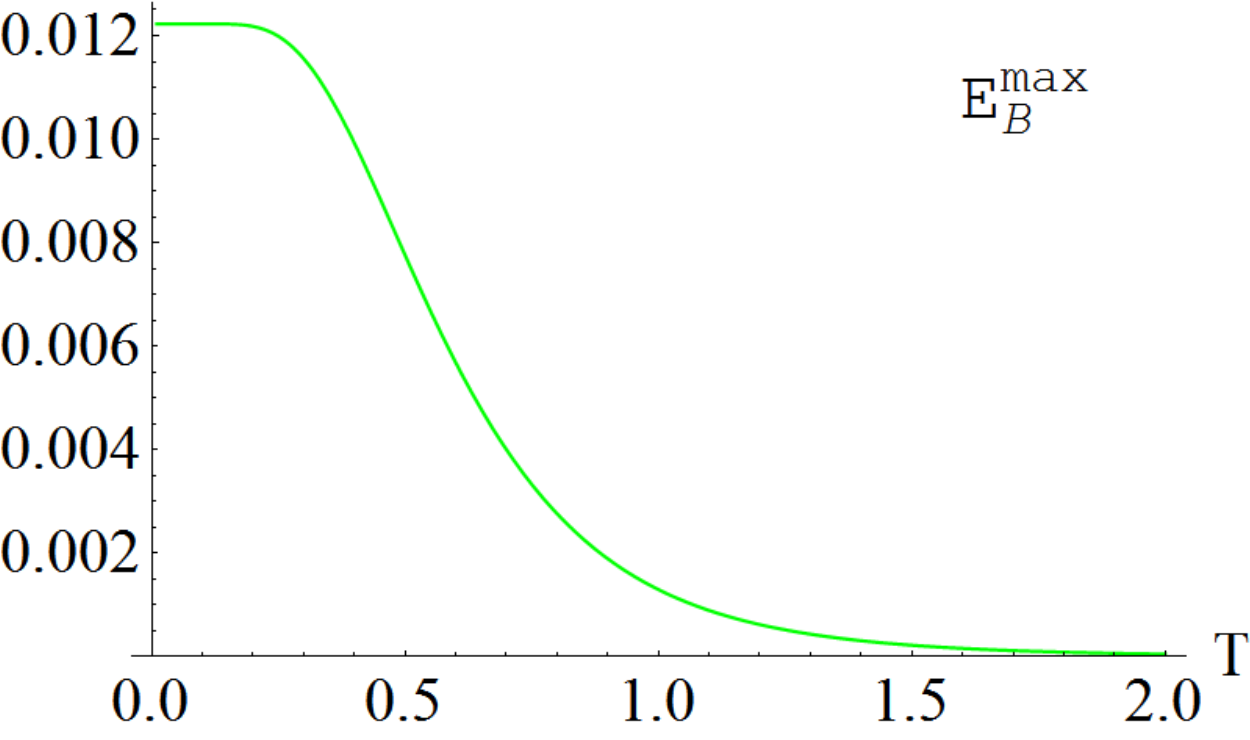}
     \includegraphics[width=4.275cm]{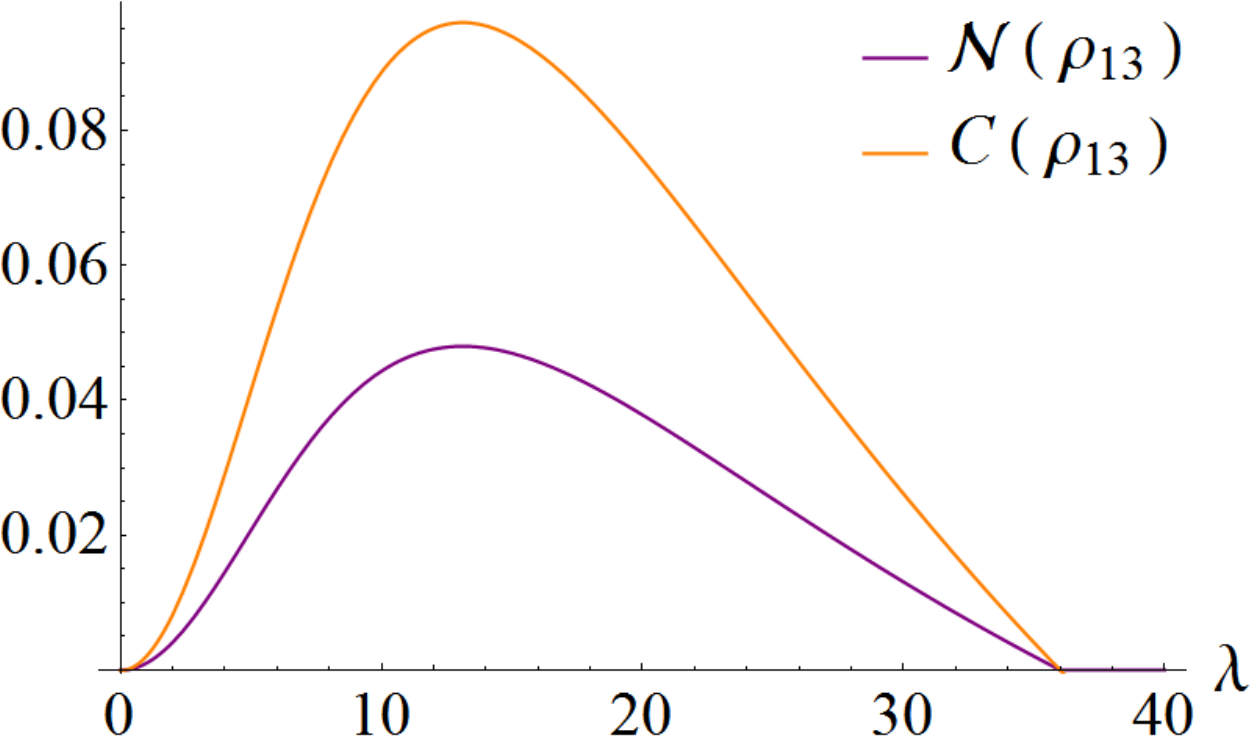}   
	\includegraphics[width=4.275cm]{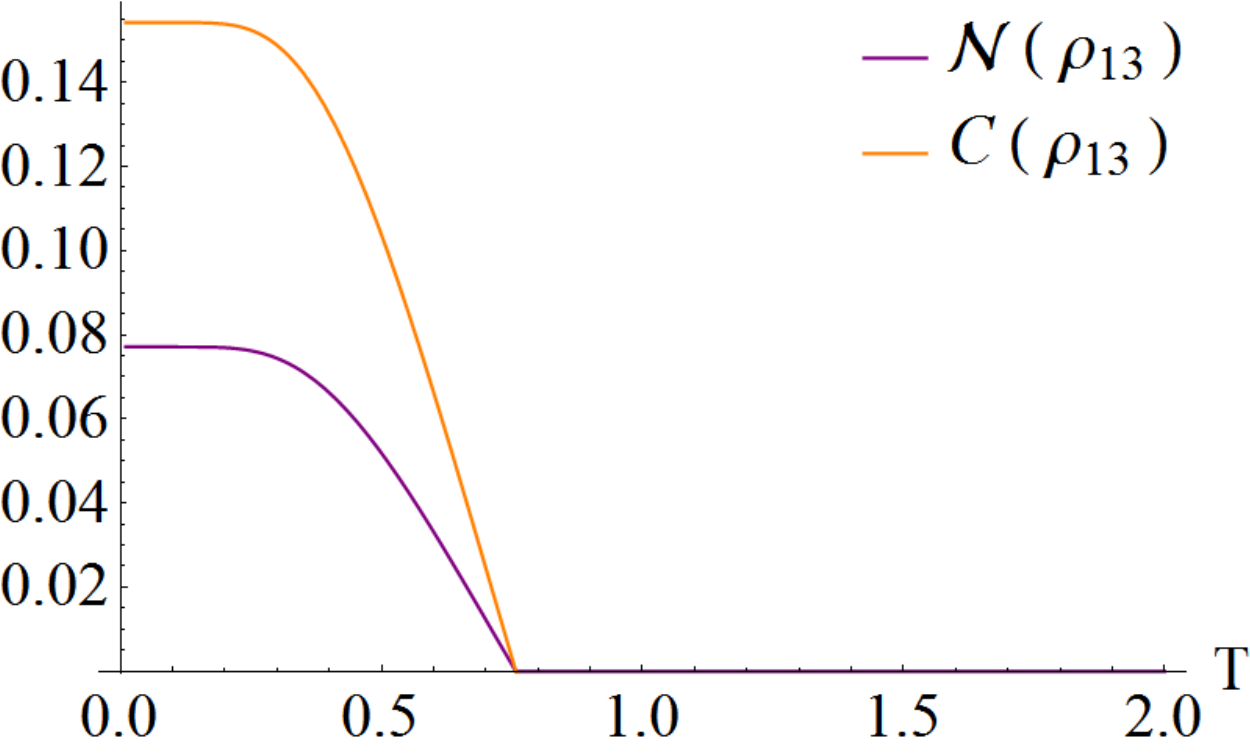}
\includegraphics[width=4.275cm]{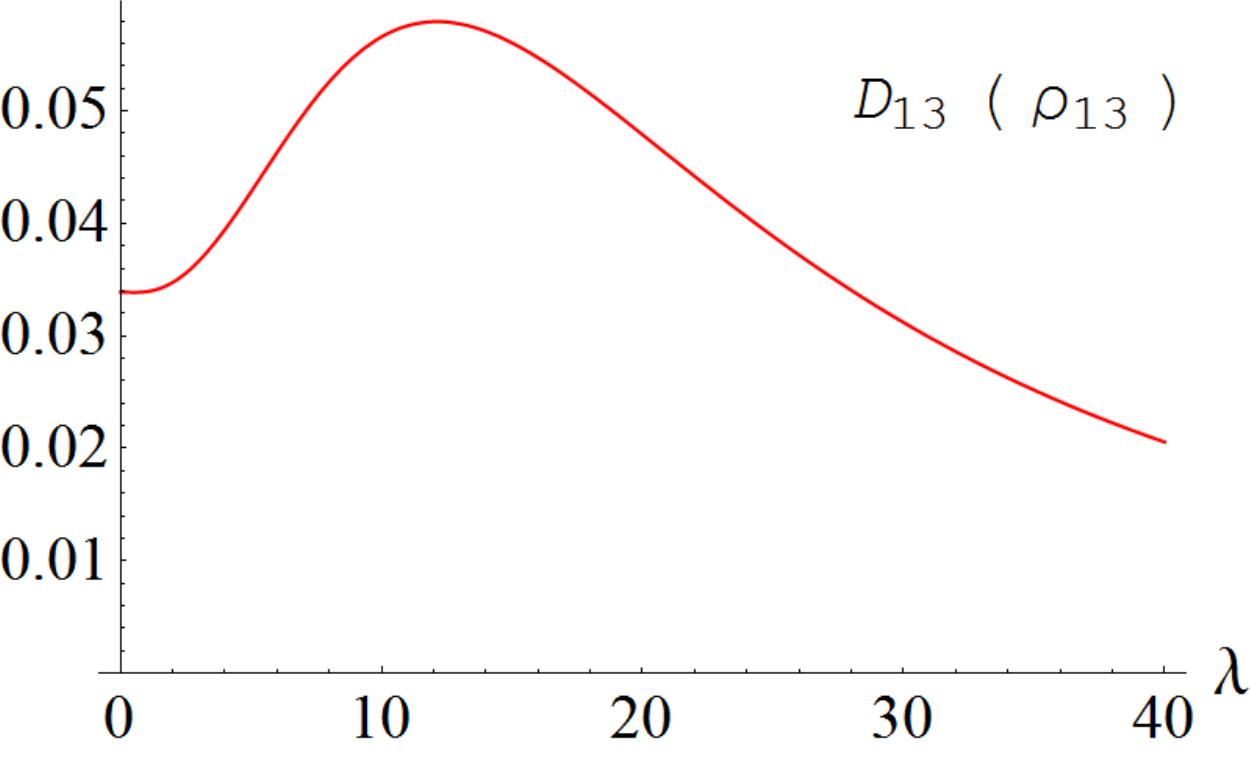}
	\includegraphics[width=4.275cm]{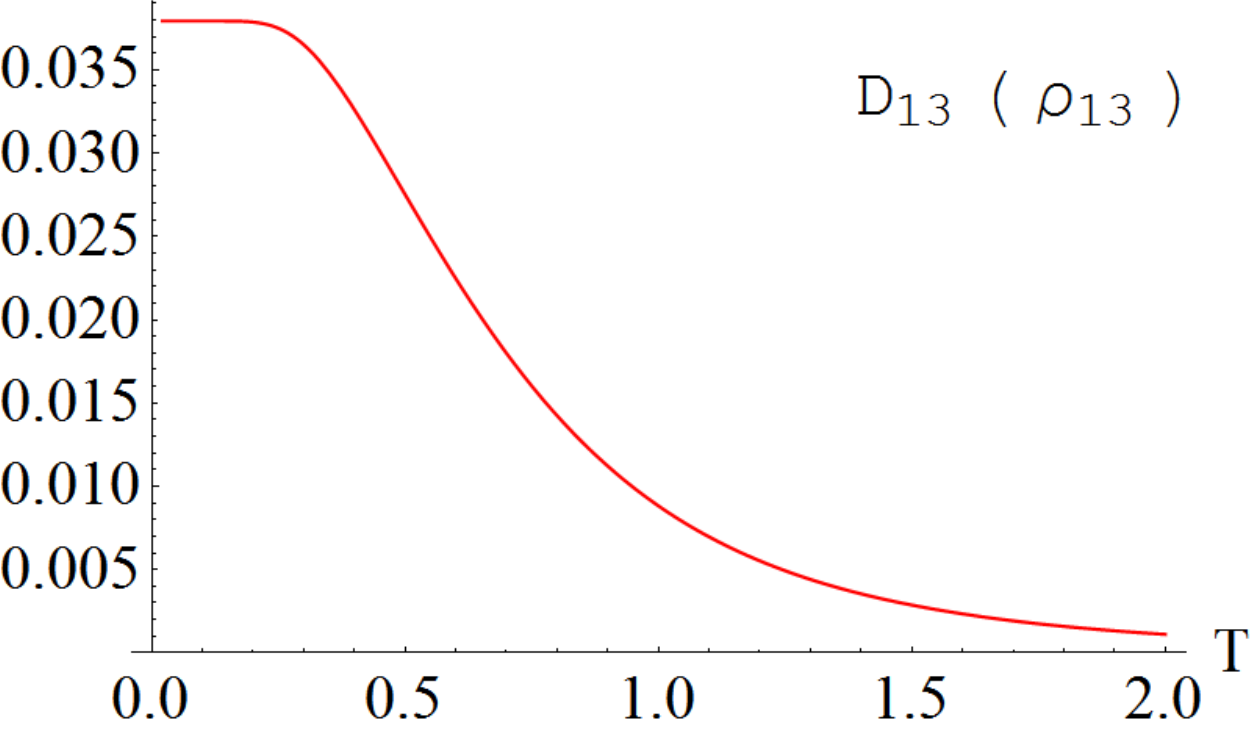}
	\caption{To the left: for \textit{T}=1 and $\kappa=5$ teleported energy $E_B^{max}$ (top), measurements of entanglement (middle) negativity $\mathcal{N}(\rho_{13})$  (purple) and concurrence $C(\rho_{13})$ (orange); quantum discord $D_{13}(\rho_{13})$ (red) (bottom); to the right: for $\lambda=10$ and $\kappa=2$ the same quantities as in the left column}
\label{new3}
\end{figure}

With the increase of the temperature, the region of the phase space ($\kappa.\lambda$) in which there is not entanglement expands from only $\lambda=0$ and $\kappa=0$ at $\textit{T}=0$ to a non trivial region whose area increases with the temperature. On the other hand, the quantum discord is different from zero in all the phase space ($\kappa.\lambda$). Similarly the teleported energy, is different from zero but with the only exception of $\lambda=0$. On the right column of figure \ref{new3} it can be seen that in the large temperature limit, for $\lambda=10$ and $\kappa=2$ the quantum dissonance, and not entanglement, is the resource of the QET protocol, similar results to the ones found on reference \cite{Frey} for the two qubit system. However, on the left column of figure \ref{new3} we can observe for \textit{T}=1 and $\kappa=5$ the same behavior of the teleported energy, measures of entanglement and correlations; as in the case of the ground state; meaning that only the quantum discord is different from zero in the limit of $\lambda=0$.  This result is remarkable since it shows that  it is possible to have regimes with non vanishing quantum dissonance, in which it is impossible to teleport energy for a finite temperature.  

The vanishing of the teleported energy $E_B^{max}$ when $\lambda=0$  lead us to make the analogy between the QET across the three-qubit open Ising chain and the Field Effect Transistor.  A Field Effect Transistor is a charge carrier device with three terminal, named: Source (S), Gate (G) and Drain (D). Energy carriers enter into the device on the Source terminal generating a current $I_S$. By manipulation of the voltage of the Gate $V_{GS}$  the current $I_D$ generated by the energy carriers leaving the device through the Drain can be modified. In the case of the model, Alice qubit will be equivalent to the Source, while Bob qubit will be equivalent to the Drain. The energy input by Alice operation $E_A$ is analogous to the current generated by the energy carriers entering into the source $I_S$, while the teleported Energy $E_B^{max}$ is to the current generated by the energy carriers leaving the drain $I_D$. The role of the voltage $V_{GS}$ will be played by the $\lambda$ parameter, representing the coupling between the spin of particle 2 and the transverse magnetic field. The limit $\lambda=0$ correspond to some value $V_{GS}^\star$ for which the output current $I_D=0$

 Since we are working with a energy dimensionless Hamiltonian, in order to fully appreciate the advantages of the QET protocol, it is necessary to calculate the efficiency $\eta$. In order to do so, it was also necessary the amount of energy that becomes the input of the protocol. Using the same optimization parameters from which the teleported energy was calculated (equation \ref{MAX}), the input energy $E_A$ can be calculated:
\begin{equation}
E_A=\mathcal{R}(\lambda,\kappa,\textit{T})-2\mathcal{B}(\lambda,\kappa,\textit{T})
\label{MAX2}
\end{equation}

\begin{figure}[h]
    \includegraphics[width=4.275cm, height=3.45cm]{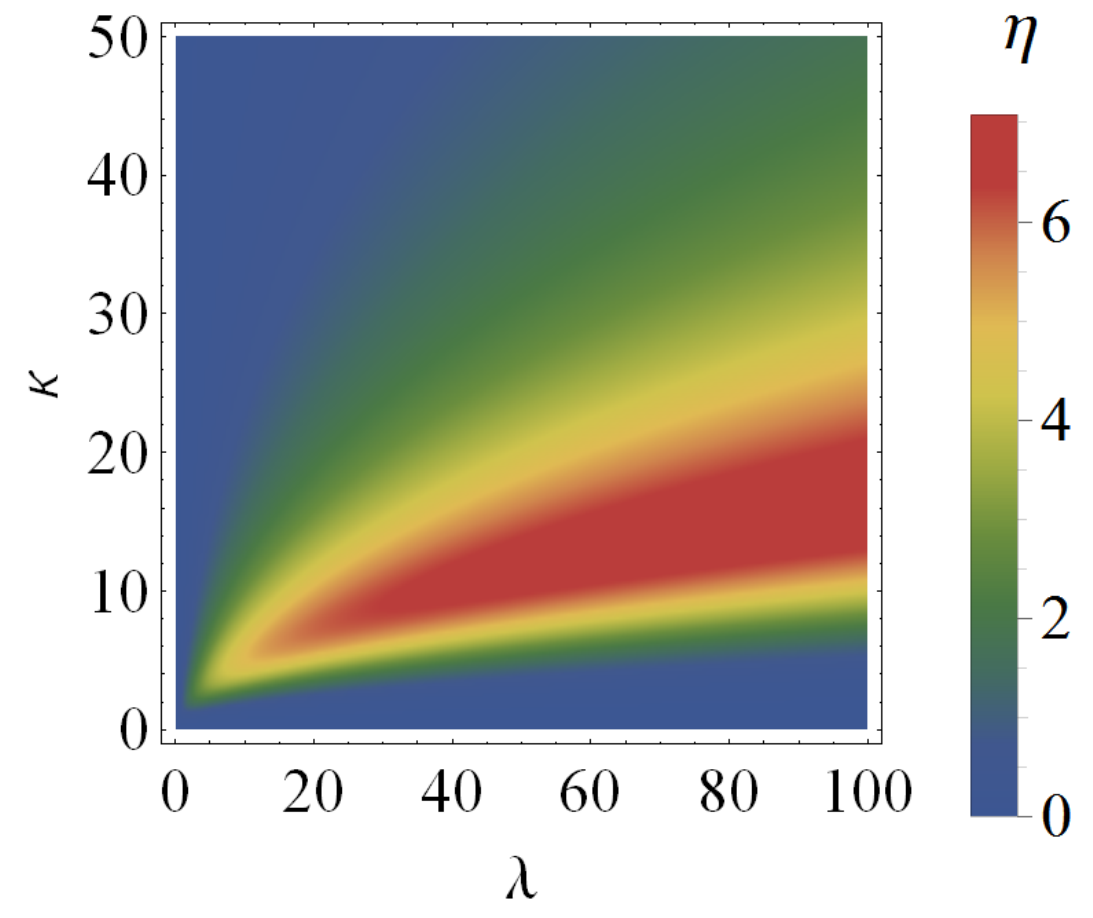}
\includegraphics[width=4.275cm]{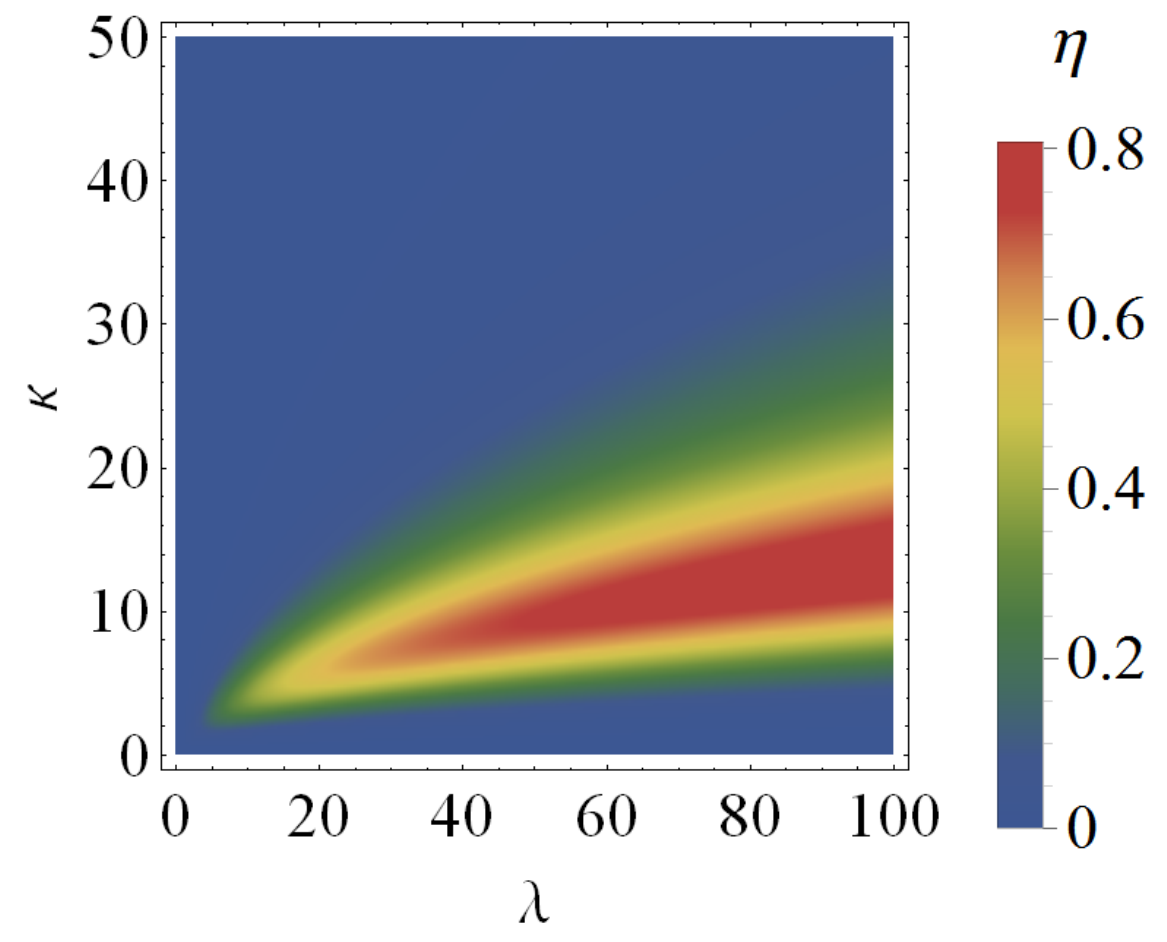}
	\caption{Efficiency of the QET protocol; $\eta=100 \ E_B^{max}/E_A$, where $E_A$ is the average energy input into the system by Alice's operation, and $E_B^{max}$ is the maximum average of teleported energy. \textit{T}=0 (left), \textit{T}=1 (right)}
\label{efficiency}
\end{figure}

The definitions of the functions $\mathcal{R}$ and $\mathcal{B}$ [\ref{RB}] make this quantity to be always positive for any value of the parameters $\lambda$ and $\kappa$ and for any temperature $\textit{T}$; a necessary condition in order to obtain energy teleportation. The efficiency $\eta$ of the QET protocol can be seen on figure \ref{efficiency} for two different temperatures \textit{T}=0, and \textit{T}=1. The maximum efficiency occurs in the case of the ground state, with a value of the order of 6\%. Similar to the teleported energy and the quantum correlations, for non-zero temperature, the maximum efficiency decreases with the increase of the temperature. The decrease of the correlations (figures \ref{compare} and \ref{compare3}) brings also the reduction on the phase space of the system in which the QET protocol can be applied efficiently. For \textit{T}=0, Gibbs states are strong local passive, in other words,  no energy extraction is possible by local operations; therefore even with an efficiency of 6$\%$, the QET protocol offers the possibility of energy extraction from the system. 

\section{\label{sec:level4} General Measurements for $\lambda=0$}

In this section we will study in more detail the behavior of the system when $\lambda=0$. Let it be $K_A(\alpha)$ the most general measurements that Alice can do to the system of the two qubits 1 and 2, while satisfying the no energy input into qubit 3; a condition like equation (\ref{restriction}). 
\begin{equation}
K_A(\alpha)=\left[a_\alpha\mathcal{I}_1+\v{b}_\alpha\cdot\gv{\sigma}_1\right] \otimes \mathcal{I}_2 +\left[c_\alpha \mathcal{I}_2+\v{d}_\alpha  \cdot \gv{\sigma}_1\right]\otimes\sigma_{2,x}
\label{genMeas}
\end{equation}
\begin{eqnarray*}
\sum_\alpha K_A^\dagger(\alpha)K_A(\alpha)=\mathcal{I}_1\otimes\mathcal{I}_2
\end{eqnarray*}

Where $a, \v{b}, c$ and $\v{d}$ are complex coefficients that depend on the measurement result $\alpha$. Since these are general measurements, the result $\alpha$ is not restricted to take the values of $\pm1$, as it was on the case of projective measurements. In fact, the $\alpha$ label in this case represents two possible results $\alpha_1$ and $\alpha_2$ that can be obtained from the measurement of qubit 1 and 2 respectively.  In addition, let us consider the most general unitary operation $U_B(\alpha)$ that Bob  can perform on qubit 3:
\begin{equation}
U_B(\alpha)=\exp{\left[-\imath \v{r}_\alpha\cdot\gv{\sigma}_3\right]}=\cos{(r_\alpha)\mathcal{I}_3-\imath \uv{r}_\alpha\cdot\gv{\sigma}_3 \sin(r_\alpha)}
\label{unitary}
\end{equation}

   Where the dependence of Alice's Measurement result $\alpha$  is contained in the real vector $\v{r}_\alpha=r_\alpha\uv{r}_\alpha$. Then, similar to equation (\ref{QET-formula}), the average energy loss of the system after the QET protocol can be written as: 
\begin{eqnarray}
E_B&=&\sum_\alpha \trace{K_A^\dagger(\alpha) K_A(\alpha) U_B^\dagger(\alpha)\commute{U_B(\alpha)}{H_B}\rho} \ \ \  
\label{QET-formula2}
		\end{eqnarray}

Where $\rho$ is the Gibbs state of the system given by equation (\ref{density}) and $H_B$ (equation \ref{BobLocal}) is the local Hamiltonian around Bob. Since Bob's operation is a unitary operation (equation \ref{unitary}), then the energy loss can be written in terms of the amplitude of oscillation $r_\alpha$ as:
\begin{eqnarray}
&& 
E_B=\sum_\alpha A_\alpha \sin{(2 r_\alpha)}-B_\alpha \sin{(r_\alpha)}^2
\\&&
A_\alpha=-\frac{\imath}{2} \trace{K^\dagger_A(\alpha)K_A(\alpha)\commute{\uv{r}_\alpha\cdot\gv{\sigma}_3}{H_B} \rho}
\\ &&
B_\alpha=-\trace{K_A^\dagger(\alpha)K_A(\alpha)\left(\uv{r}_\alpha\cdot\gv{\sigma}_3 \right)\commute{\uv{r}_\alpha \cdot \sigma_3}{H_B} \rho}
\end{eqnarray}

By explicit calculation of $A_\alpha$, without specifying the $\alpha$ dependence of the coefficients $a, \v{b}, c, \v{d}$ and the components of the unit vector $\uv{r}$, it can be shown that $A_\alpha=0$ on the limit when $\lambda=0$. To prove this result it is necessary to use the equations ($\ref{lambda01}-\ref{lambda03}$) that can be found on the appendix A.  Therefore in this limit the average energy loss can be written as:
\begin{equation}
E_B=-\sum_\alpha B_\alpha \sin{(r_\alpha)}^2 \ \ \ \ \ \ \ \lambda=0
\end{equation}

As can be seen from the previous equation the sign of the energy loss on the limit $\lambda=0$ depends only on the coefficient $B_\alpha$. If this one is always positive, meaning that the system gains energy instead of losing it, then energy teleportation is not possible. Let us define from the coefficients of the general measurement $K_A(\alpha)$ on equation (\ref{genMeas}) the complex column vector $\v{V}=\left(a, b_1, b_2, b_3, c, d_1, d_2, d_3\right)^t$   where the $t$ represents the transposition operation and the $\alpha$ dependence has been omitted for simplification of the notation. Similarly, let us write the unitary real vector $\uv{r}_\alpha$ as a column vector $\uv{R}=\left(\hat{r}_1,\hat{r}_2,\hat{r}_3\right)^t$; where again the $\alpha$ dependence has been omitted. Then let us define the matrices of operators $\mathcal{M}_{\beta \gamma}$ and $\mathcal{O}_{j k}$, independent of $\alpha$, with the  indexes $\beta$ and $\gamma$ from $\{1,2, ... 8\}$ and $j k$ from $\{1,2,3\}$ such as:
\begin{eqnarray}
&&
K_A^\dagger(\alpha) K_A(\alpha)= \v{V}^\dagger \mathcal{M} \v{V}
\\ &&
\mathcal{O}_{j k}=\sigma_{3, j} \commute{\sigma_{3, k}}{H_B}
\end{eqnarray}  

Then the coefficient $B_\alpha$ can be written as
\begin{equation}
B_\alpha=-(\v{V}^*_\beta \uv{R}_j) \trace{\mathcal{M}_{\beta \gamma} \mathcal{O}_{j k} \rho}(\uv{R}_k \v{V}_\gamma)
\end{equation} 

Where the sum over repeated indexes is implied. Since the vectors $\uv{R}$ and $\v{V}$ depend on the measurement result $\alpha$; then let us assume that it is possible to chose them such as $\uv{R}\otimes\v{V}$ is an eigenvalue of the $24 \times 24$ matrix $\mathcal{P}$ with elements given by  $\mathcal{P}_{\beta \gamma j k}= -\trace{\mathcal{M}_{\beta \gamma} \mathcal{O}_{j k} \rho}$. Then to study the positivity of the coefficient $B_\alpha$ it is necessary to calculate the eigenvalues of $\mathcal{P}$.
\begin{eqnarray}
&&
\mathcal{P}=\begin{pmatrix}
p_ 0 & 0 &\kappa p_1 \\
0 & (1+\kappa^2) p_0 & 0 \\
\kappa p_1 & 0 & \kappa^2 p_0
\end{pmatrix}
\end{eqnarray}

Where the matrices $p_0$ and $p_1$ (\ref{p1}) are $8\times8$ complex matrices. The eigenvalues of $\mathcal{P}$ are eight times degenerated each, and they are given by:
\begin{equation}
\text{Eig}\left(\mathcal{P}\right)=\Big\{0, (1+\kappa^2)\left(\mathcal{C}_1\pm \mathcal{C}_2 \sqrt{1+k^2}\right)\Big\}
\end{equation}

Where $\mathcal{C}_1$ (\ref{c1}) and $\mathcal{C}_2$ (\ref{c2}) are two real positive functions with $\mathcal{C}_1-\mathcal{C}_2\sqrt{1+\kappa^2}>0$. Therefore, all the eigenvalues are positive, which implies that it is impossible to teleport energy, not even with General Measurements. This is a remarkable result, since at the limit $\lambda=0$ there are quantum correlations that could   act as the resource for the QET.

	\section{\label{sec:level5} Conclusions }

We applied the QET protocol to a three spin open chain model, from one edge spin to the other edge spin.  We calculted an optimal QET protocol using projective measurements, defined such as the maximum amount of teleported energy is obtained, for the ground state and finite temperature case. For this optimal QET protocol, we obtained an efficiency of the order of  6$\%$ in the region in which it is not possible any energy extraction by local operations due to the strong local passivity of the Gibbs states for \textit{T}=0. As opposed to local energy extraction by general operations, the QET protocol allows to extract energy for every value of \textit{T}, $\kappa$ and $\lambda$ with the exception of $\lambda=0$. 

 For low temperature regimes, negativity and concurrence are well correlated with the amount of optimal teleported energy and can be regarded as QET resource. However, for high temperature regimes although the negativity and concurrence vanish at some critical temperature; energy teleportation is possible. In this case, non-entanglement resource like quantum dissonance yields high-temperature QET as in the two qubit model. In addition, we found that the negativity and the concurrence between the edge spins are exactly zero in the case $\lambda=0$. Even further, we proved that employing the most general operations to the system it is impossible to teleport energy in the case of $\lambda=0$ and finite temperature. Even though there are quantum correlations, different from entanglement, no energy can be teleported, in contrast with the regime of high temperatures and $\lambda\neq0$ in which it is possible to teleport energy due to the quantum dissonance. 

	\appendix
	\section{Eigenvalues and Density Matrix for the Open Chain}
		\label{Appendix}
		\ \ \ \ The spectrum of the Hamiltonian in equation (\ref{Hamiltonian}) is composed by eight eigenvalues:
		\begin{eqnarray}
			&& 
			E_0=-E_7=-\frac{1}{3}[\lambda+x_0]
			\\ &&
			E_1=-E_6=-E_0-\frac{1}{2}\left(x_0+\frac{y_0}{\sqrt{3}}\right)
			\\&&
			E_2=-E_5=-\lambda
			\\ &&
			E_3=-E_4=-E_0-\frac{1}{2}\left(x_0-\frac{y_0}{\sqrt{3}}\right)
			\end{eqnarray}
		Where it was defined:
			\begin{eqnarray*}
	&&x_0=r_0 \cos\left(\frac{\theta}{3} \right) \ \ \ \ 		y_0=r_0 \sin\left(\frac{\theta}{3}\right)
 \\ &&
		r_0=4\sqrt{3+3\kappa^2+\lambda^2}
	\\&&
g(\lambda, \kappa)=\lambda^2     (\kappa^2+20)+4(3+3\kappa^2+\kappa^4)
\\ &&	\tan\left(\theta\right)=\frac{\sqrt{27\left[4(\lambda^2-1)^2+\kappa^2g(\lambda, \kappa)\right]}}{\lambda\left[16-9\kappa^2-2(\lambda^2-1)\right]}		
			\end{eqnarray*}
	
		The Eigenvalues are symmetric with the exchange of $\kappa$ with $-\kappa$.  For $\lambda>0$ the lowest eigenvalue is $E_0$, on the other hand  for $\lambda<0$ the lowest eigenvalue is $E_1$. For $\lambda=0$ the system described by equation (\ref{Hamiltonian}) is completely degenerated, and the  energies are 
\begin{eqnarray}
			&& 
			E_0=E_1=-E_7=-E_6=-2\sqrt{1+\kappa^2}
			\\ &&
           	E_2=-E_5=E_3=-E_4=0
\label{degeneracy}
			\end{eqnarray}

This degeneracy of the ground state only occurs on the physical limit of $\lambda=0$. For any value of $\lambda$; the Eigenvalues $E_2$ and $E_5$ the corresponding Eigenvectors are:
	 \begin{eqnarray}
	&&		\ket{E_{2}}=\frac{1}{\sqrt{2}}\left[-\ket{011}+\ket{110}\right]
			\\ &&
			\ket{E_{5}}=\frac{1}{\sqrt{2}}\left[-\ket{001}+\ket{100}\right]
	\end{eqnarray}
	
	The  eigenvectors $\ket{E_{AC}}$ are associated to the Eigenenergies E=$\left\{E_1, E_3, E_7\right\}$. 
		\begin{eqnarray}
&& (E+\lambda)(E-2-\lambda)(E+2-\lambda)=4 \kappa^2(E-\lambda)
\\
		&&	\ket{E_{AC}}=\frac{1}{N_{AC}}\left[A\ket{000}+\ket{011}+C\ket{101}+\ket{110}\right] \ \ \ \ \ \ 
\end{eqnarray}

	Where the Normalization constant $N_{AC}$ and the probabilities amplitudes A, C are functions of the associated Eigenvalue E.
\begin{eqnarray*}
 &&
		A=\frac{2\kappa}{E-2-\lambda} \ \ \ \ \  C=\frac{2\kappa}{E+2-\lambda} \\ &&
h(\lambda,E)=8(\lambda^2-1)(E-\lambda-2)(E-\lambda+2)
\\ &&
i(\lambda, E)=16\kappa^2\left[(E-2\lambda)(E-\lambda)+2\right]
\\ &&
P_{AC}=\frac{1}{h(\lambda, E)+i(\lambda, E)}
\\ &&
 	\frac{1}{N_{AC}}=\sqrt{P_{AC} (E-\lambda+2)^2(E-\lambda-2)^2}
	\end{eqnarray*}

	On the other hand, the eigenvectors $\ket{E_{DF}}$ are associated to the Eigenenergies $E=\left\{E_0, E_4, E_6\right\}$, solutions of the same equation as the Eigenvalues associated to  $\ket{E_{AC}}$ with the exchange of $\lambda$ with $-\lambda$. 
	\begin{eqnarray}
&&
(E-\lambda)(E-2+\lambda)(E+2+\lambda)=4 \kappa^2(E+\lambda)
\\	&&
		\ket{E_{DF}}=\frac{1}{N_{DF}}\left[\ket{001}+F\ket{010}+\ket{100}+D\ket{111}\right] \ \ \ \ \ \ \ \ \ 
\end{eqnarray}
\begin{eqnarray*}
 &&	
F(\lambda)=A(-\lambda)  \ \ \ D(\lambda)=C(-\lambda) \  \ \ \ 
	 \frac{1}{N_{DF}(\lambda)}=\frac{1}{N_{AC}(-\lambda)}
			\end{eqnarray*}


In particular for $\lambda=0$ due to the degeneracy of the system (\ref{degeneracy}) it is possible to rewrite the Eigenstate for the ground state $\ket{g}$ as a linear combination of $\ket{E_{AC}}$ with energy $E=E_1$ and $\ket{E_{DF}}$ with energy $E=E_0$:
\begin{equation}
\ket{g}=a \ket{E_{AC}}+ b \ket{E_{DF}}
\end{equation}

Where one of the coefficients a and b will be determined by the normalization condition. This is possible since on this limit $E_0=E_1=-2\sqrt{1+\kappa^2}$, and a linear combinations of eigenstates of a Hamiltonian with the same Eigenvalue is also an Eigenstate of the Hamiltonian. Considering that the Hamiltonian (\ref{Hamiltonian}) commutes with the Pauli operator $\sigma_{2,x}$ it is possible to have a common base of Eigenstates for both operators; then after a few calculations it is possible to prove that:
\begin{eqnarray}
&&\ket{g}=\ket{\phi}_1\ket{\psi}_2\ket{\phi}_3
\label{ground}
\\ &&
\ket{\phi}_1=\alpha \ket{0}_1+\beta \ket{1}_1
\\ &&
\ket{\psi}_2=\frac{1}{\sqrt{2}}\left(\ket{0}_2\pm\ket{1}_2\right)
\\ &&
\ket{\phi}_3=\gamma \ket{0}_3+\delta \ket{1}_3
\end{eqnarray}

Where the subsidences after the kets are related to the particle label. The probabilities amplitudes, in terms of only D and F evaluated for $\lambda=0$ and $E=E_0$, are given by:
\begin{eqnarray*}
&&\abs{\alpha}^2=\frac{1}{1+D^2} \ \ \ \ \ \ \ \ \ \ \ \ \  \ \ \ \ 
 \abs{\beta}^2=\frac{D^2}{1+D^2} 
\\ &&
\abs{\gamma}^2=\frac{1+D^2}{D^2(F^2+D^2+2)}  \ \ \ \ 
\abs{\delta}^2=\frac{1+D^2}{F^2+D_2+2} 
\end{eqnarray*}

In other words the ground state of the system can be written as a totally separable state in the limit of $\lambda=0$.  If the ground state is given by equation (\ref{ground}), then there is not entanglement or classical correlations  between qubit 1 and 3. This is possible since this ground state was constructed as a linear superpositions of ground states, with the correct coefficients a and b.  However, assuming that it is possible to achieve the limit \textit{T}=0 and $\lambda=0$, in the case that both parameters are gradually decreased, the ground state of the system will be given by $\ket{E_{DF}}$ with $E=E_0=-2\sqrt{1+\kappa^2}$ for $\lambda=0$, a non separable state. Similar calculations are possible for the remaining six eigenstates associated to the Energies $E={0,2\sqrt{1+\kappa^2}}$. Therefore, since the system is degenerated, the lack or presence of correlations between the spins 1 and 3, for the case $\lambda=0$ will depend on  the state in which the system chose to be. 
 
	If the system of the three particles is weakly coupled to a thermal bath at temperature \textit{T}; then the probability that the system has an energy E will be given as:  
			\begin{equation}
			p_j(\textit{T})=\frac{1}{Z} \exp{\left(-\frac{E_j}{\textit{T}}\right)}
			\label{canonical}
			\end{equation}

			Where the Temperature \textit{T}, similar to the energies is a dimensionless parameter, and Z is the partition function in the canonical ensemble. The state of the system of the three particles in thermal equilibrium at temperature \textit{T},  on the computational basis $\left\{\ket{000}, \ket{001}, ...., \ket{111}\right\}$ and the partition function Z are given by:
\begin{widetext}		
	\begin{eqnarray}
			Z&=&\sum_{j=0}^{7} \exp{\left(-\frac{E_j}{\textit{T}}\right)}=2\left[\cosh{\left(\frac{E_0}{\textit{T}}\right)}+\cosh{\left(\frac{E_6}{\textit{T}}\right)}+\cosh{\left(\frac{E_5}{\textit{T}}\right)}+\cosh{\left(\frac{E_4}{\textit{T}}\right)}\right]
\\
\rho(\textit{T})&=&\sum_{j=0}^{7}p_j(\textit{T})\ketbra{E_j}{E_j}=\frac{1}{2}
			\begin{pmatrix}
			F_1 & 0 &  0 & F_2&  0 & F_4& F_2& 0\\
			0 & J_3 +p_5 & J_2 &0 & J_3 -p_5&  0& 0 & J_5 \\ 
			0 & J_2 & J_1 &  0 & J_2 & 0 & 0 & J_4\\ 
			F_2 & 0 & 0 & F_3 + p_2 & 0 & F_5& F_3 - p_2 & 0 \\
			0 & J_3 - p_5 & J_2 & 0 & J_3 + p_5 & 0 & 0& J_5\\
			F_4& 0& 0& F_5& 0& F_6& F_5& 0\\ 
			F_2 & 0& 0& F_3 - p_2 & 0 & F_5& F_3 + p_2& 0\\
			0 & J_5& J_4 & 0 & J_5& 0& 0& J_6\\
			\end{pmatrix}
			\label{density}
			\end{eqnarray}
\end{widetext}

			Where the $F_i=F[f_i]$ and $J_i=J[j_i]$ for i=$\{ 1,2,3,4,5,6 \}$ are functionals of $f_i(E)$ and $j_i(E)$ functions, that are associated to the group of Eigenenergies $\{E_1, E_3, E_7\}$ and $\{E_0, E_4, E_6\}$ respectively. The functional, which will also depend on the probabilities $p_j$ of the canonical ensemble (\ref{canonical}), are defined as follows:
			\begin{eqnarray}
			&&
			F[f_i]=F_i=2 \sum_{k=1,3,7} f_i(E_k) p_k(\textit{T})
			\\ &&
			J[j_i]=J_i=2 \sum_{k=0,4,6} j_i(E_k) p_k(\textit{T})
		     \label{functionals}	
		\end{eqnarray}

			The functions $f_i$ and $j_i$ are build with polynomials of the Eigenvalue E. In order to obtain the most simple representation of those functions, the characteristic equation for each set of Eigenvalues  ($\{E_1, E_3, E_7\}$ and $\{E_2, E_4, E_6\}$) was used. The functions were found to be as follows:
	
	\begin{eqnarray*}
		&&
			f_1(E)=P_{AC} \  4\kappa^2(E+2-\lambda)^2
			\\ &&
			f_2(E)=P_{AC} \ 2\kappa(E+2-\lambda)^2(E-2-\lambda)
			\\ &&
			f_3(E)=P_{AC} \ (E+2-\lambda)^2(E-2-\lambda)^2
			\\ &&
			f_4(E)=P_{AC} \ 4\kappa^2(E+2-\lambda)(E-2-\lambda)
			\\ &&
			f_5(E)=P_{AC} \ 2\kappa (E-2-\lambda)^2(E+2-\lambda)
			\\ &&
			f_6(E)=P_{AC} \ 4\kappa^2(E-2-\lambda)^2
              \\ &&
              j_k(E)=f_k(E) \ \ \  \text{with $\lambda\rightarrow-\lambda$ for all k=1,2,...6}
			\end{eqnarray*}
					
			Since the Eigenergies $E_0, E_4, E_6$ are obtained from $E_7, E_3, E_1$ with the change of $\lambda \rightarrow -\lambda$, and the same for the functions $j_k(E)$ from $f_k(E)$; then any expression containing both functional on the form $F_k\pm J_k$ for any k=1,2,...,6 is symmetric under the exchange  $\lambda $ with $-\lambda$.  On the other hand, with respect to the coupling parameter $\kappa$: $F_2, F_5, J_2, J_5$ are odd functions, while the rest are even functions. For all values of $\lambda, \kappa$ and \textit{T} the functions $F_i$ and $J_i$ satisfy 
	\begin{eqnarray*}
	2F_4+\kappa (F_5-F_2) &=& 0
	\\
	2 J_4+\kappa (J_5-J_2)&=&0
\\ 
	\frac{1}{2}\left(F_1+F_6+J_1+J_6\right)+F_3+J_3+&p_2&+p_5=\sum_{j=0}^{7}p_j=1
	\end{eqnarray*}
   

With the definitions of the F and J functionals, the following even functions with respect to both parameters $\kappa$ and $\lambda$ were defined 
 \begin{eqnarray}
	 &&
	\mathcal{A}(\lambda, \kappa,\textit{T})=F_4-F_3-J_3+J_4+p_2+p_5
	\\ &&
	\mathcal{R}(\lambda, \kappa,\textit{T})=-\frac{1}{2}\left(F_1+J_1-F_6-J_6\right)
	\\ &&
	\mathcal{B}(\lambda, \kappa,\textit{T})=\frac{\kappa}{2}\left(F_2+F_5+J_2+J_5\right)
	\label{RB}
	\end{eqnarray}


Through numerical evaluation and analytical calculations it was verified that $0\leq R(\lambda, \kappa, \textit{T}) \leq 1$ and $B(\lambda,\kappa, \textit{T})<0$ for all values of the coupling parameter $\kappa$, $\lambda$ and finite temperature \textit{T}.  In particular for $\lambda=0$, as can be done by analytical calculation, the function $\mathcal{A}(0, \kappa, T)=0$ since the functionals satisfy:
		\begin{eqnarray}
	&& 
	F_4-F_3+p_2=0 \ \ \ \ \ \ \ \ \ \ \ \ \  \ \ \ \lambda=0
\label{lambda01}
\\ &&
	2(F_2+F_5)=\kappa (F_1-F_6) \ \ \ \ \ \  \lambda=0
\\ &&
   2 F_5=\kappa \left(2 p_2+F_4-F_6\right) \ \ \ \ \ \ \lambda=0
\label{lambda03}
\end{eqnarray}

In this limit of $\lambda=0$ it is convenient to define also the functions $\mathcal{C}_1=F_6-F_1$ and  $\mathcal{C}_2=F_1-2F_3+F_6-2 p_2$; which can be written as:
\begin{eqnarray}
&& 
\mathcal{C}_1=\frac{4}{Z\sqrt{1+\kappa^2}}\sinh{\left(\frac{2\sqrt{1+\kappa^2}}{T}\right)}
\label{c1}
\\ &&
\mathcal{C}_2=\frac{4}{Z(1+\kappa^2)^2}\left[\cosh{\left(\frac{2\sqrt{1+\kappa^2}}{T}\right)}-1\right]
\label{c2}
\end{eqnarray}

As it can be seen both functions are always positive for any value of the temperature $\textit{T}$ and the coupling parameter $\kappa$. In addition, they satisfy $\mathcal{C}_1\geq\sqrt{1+\kappa^2} \mathcal{C}_2$. To end this appendix we will define the two matrices that were build with the previous functions:
\begin{widetext}
	\begin{eqnarray*}
&& p_0=
	\begin{pmatrix}
\mathcal{C}_1 &  0 & 0 & -\mathcal{C}_2 & 0 & -\mathcal{C}_2 k & 0& 0\\
0 & \mathcal{C}_1& -\imath \mathcal{C}_2 &  0 & -\mathcal{C}_2 k &  0 & 0& 0\\ 
0 & \imath \mathcal{C}_2 & \mathcal{C}_1 &  0&  0&  0& 0& -\imath\mathcal{C}_2 k \\ 
-\mathcal{C}_2 & 0 & 0& \mathcal{C}_1& 0& 0& \imath\mathcal{C}_2 k&0\\ 
 0 & -\mathcal{C}_2 k & 0& 0 & \mathcal{C}_1& 0& 0& -\mathcal{C}_2\\
-\mathcal{C}_2 k & 0& 0& 0& 0& \mathcal{C}_1& -\imath \mathcal{C}_2& 0\\
0& 0& 0& -\imath\mathcal{C}_2 k & 0 & \imath\mathcal{C}_2& \mathcal{C}_1& 0\\
0& 0& \imath \mathcal{C}_2 k& 0& -\mathcal{C}_2& 0& 0& \mathcal{C}_1\\
\end{pmatrix}
\label{po}
 \ \  p_1=
\begin{pmatrix}
0 & \mathcal{C}_2 k& 0& 0& \mathcal{C}_1& 0& 0& \mathcal{C}_2\\
\mathcal{C}_2 k& 0& 0& 0& 0& \mathcal{C}_1& \imath \mathcal{C}_2& 0\\
0 & 0& 0& \imath \mathcal{C}_2 k& 0& -\imath \mathcal{C}_2& \mathcal{C}_1k& 0\\
0 & 0& -\imath\mathcal{C}_2 k& 0& \mathcal{C}_2 & 0& 0& \mathcal{C}_1\\
\mathcal{C}_1 & 0& 0& \mathcal{C}_2 & 0& \mathcal{C}_2 k& 0& 0\\
0& \mathcal{C}_1& \imath \mathcal{C}_2& 0& \mathcal{C}_2 k& 0& 0& 0\\
0& -\imath\mathcal{C}_2& \mathcal{C}_1 & 0& 0& 0& 0& \imath \mathcal{C}_2 k\\
\mathcal{C}_2& 0& 0& \mathcal{C}_1 & 0& 0& -\imath \mathcal{C}_2 k& 0\\
\end{pmatrix}
\label{p1}
\end{eqnarray*}
\end{widetext}


\section{Measures of Quantum Correlations of the reduced system $\rho_{13}$}
\label{Appendix3}

In this appendix it will be explained the calculation of several quantum correlations for the reduced system $\rho_{13}$. In particular: the Concurrence, the Negativity and the Discord will be calculated. The state of the reduced system $\rho_{13}$ is obtained by taking the trace over the qubit 2 of the state of the system $\rho$  

\begin{widetext}
\begin{equation}
\rho_{13}=\mathrm{Tr}_2\left[\rho\right]=\frac{1}{4}	\begin{pmatrix}
			1+r+s+c_3& 0&0& c_1-c_2\\
			0&1+r-s-c_3& c_1+c_2 & 0 \\ 
			0&c_1+c_2 & 1-r+s-c_3 & 0\\ 
			c_1-c_2&0&0 &1-r-s+c_3 \\
			\end{pmatrix}
\end{equation}
\end{widetext}
\begin{eqnarray*}
&& r= s= \frac{1}{2}\left(F_1+J_1-F_6-J_6\right)=- \mathcal{R}(\lambda, \kappa, \textit{T})
\\ && 
c_1=\left(F_4+J_4+F_3+J_3-p_2-p_5\right)
\\ && 
c_2= \left(F_3+J_3-p_2-p_5-F_4-J_4\right)= -\mathcal{A}(\lambda, \kappa, \textit{T})
\\ && 
c_3= \frac{1}{2}\left(F_1+J_1+F_6+J_6\right)-\left(F_3+J_3+p_2+p_5\right)
\end{eqnarray*}

The labeling of r, s, $c_1, c_2, c_3$ where chosen in order to follow reference \cite{Xstates}, where the quantum correlations for two qubits in an X state was studied. The Eigenvalues of the state $\rho_{13}$, to be called $\Lambda_j$,  not to confuse with the $\lambda$ parameter of the three qubit model, can be calculated in terms of the new labeling as:
\begin{eqnarray*}
&& \Lambda_{1}=\frac{1}{4}\left(1-c_3-\sqrt{(c_1+c_2)^2+(r-s)^2}\right)
\\ 
&& \Lambda_{2}=\frac{1}{4}\left(1-c_3+\sqrt{(c_1+c_2)^2+(r-s)^2}\right)
\\ 
&& \Lambda_{3}=\frac{1}{4}\left(1+c_3-\sqrt{(c_1-c_2)^2+(r+s)^2}\right)
\\ 
&& \Lambda_{4}=\frac{1}{4}\left(1+c_3+\sqrt{(c_1-c_2)^2+(r+s)^2}\right)
\end{eqnarray*}

 The concurrence $C(\rho_{13})$, is a measure of quantum entanglement \cite{Concurrence}, that can be calculated in terms of the eigenvalues $\Lambda_C$ of $\rho_{13} {\tilde \rho_{13}}$, where $\tilde\rho_{13}=\left(\sigma_1^\mathcal{Y}\otimes\sigma_3^ \mathcal{Y}\right)\rho_{13}^\star\left(\sigma_1^\mathcal{Y}\otimes \sigma_3^\mathcal{Y}\right)$. Those eigenvalues $\Lambda_C$ are given by:
\begin{eqnarray*}
&& \Lambda_{1,C}=\frac{1}{16}\left(c_1-c_2-\sqrt{(1+c_3)^2-(r+s)^2}\right)^2
\\ &&
\Lambda_{2,C}=\frac{1}{16}\left(c_1-c_2+\sqrt{(1+c_3)^2-(r+s)^2}\right)^2
\\ &&
\Lambda_{3,C}=\frac{1}{16}\left(c_1+c_2-\sqrt{(1-c_3)^2-(r-s)^2}\right)^2
\\ &&
\Lambda_{4,C}=\frac{1}{16}\left(c_1+c_2+\sqrt{(1-c_3)^2-(r-s)^2}\right)^2
\end{eqnarray*}

 \begin{eqnarray*}
C\left(\rho_{13}\right)=  \text{max}\left\{2\text{max}\left\{\sqrt{\Lambda_{1,C}}, \sqrt{\Lambda_{2,C}}, \sqrt{\Lambda_{3,C}}, \sqrt{\Lambda_{4,C}} \right\}\right. \\
\left. -\sqrt{\Lambda_{1,C}}-\sqrt{\Lambda_{2,C}}-\sqrt{\Lambda_{3,C}}- \sqrt{\Lambda_{4,C}}, 0 \right\}
\end{eqnarray*}

With the definitions of $r, s, c_1, c_2, c_3$ and trough computational calculations  the Concurrence $C(\rho_{13})$ can be written as 
\begin{equation}
C(\rho_{13})= \text{max} \left\{\sqrt{\Lambda_{2,C}}-\sqrt{\Lambda_{1,C}}-\sqrt{\Lambda_{3,C}}- \sqrt{\Lambda_{4,C}}, 0 \right\}
\end{equation}

Another simple to calculate entanglement measure is the Negativity $\mathcal{N}(\rho_{13})$ \cite{Negativity}. This one is defined as the absolute sum of the negative eigenvalues of the partial transpose of the density matrix $\rho_{13}$ with respect to qubit 1. 
\begin{equation}
\mathcal{N}(\rho_{13})=\sum_i \frac{\abs{\Lambda_{i,N}}-\Lambda_{i,N}}{2}
\end{equation}

Where the eigenvalues $\Lambda_{i,N}$ are the eigenvalues of $\rho_{13}^{(T_1)}$. Due to the symmetry of the X states, the partial transpose with respect to qubit 1 of the state $\rho_{13}$ is obtained by the exchange of the elements (a) on the antidiagonal: $a_{14}$ with  $\ a_{32}$, and $a_{23}$ with $\ a_{41}$; which is equivalent to exchange $c_2$ with $\ -c_2$. 
\begin{eqnarray*}
&&
\rho_{13}^{(T_1)}=\rho_{13}   \ \ \ \text{with} \ \ c_2\longrightarrow -c_2
\\ &&
\Lambda_{i,N}=\Lambda_{i} \ \ \ \text{with} \ \ c_2 \longrightarrow -c_2 \ \ \ \ \text{for i=1,2,3,4}
\end{eqnarray*}

By taking the limit $\lambda=0$ it is possible to prove analytically that the eigenvalues $\Lambda_{i,N}$ are all positive for every value of the coupling parameter $\kappa$ and the Temperature T; therefore there is not Entanglement between qubits 1 and 3.  

The Quantum Discord \cite{Discord} \cite{Discord2} is a measure of the non classical correlations between two subsystems of a quantum system. The Discord include all the quantum correlations, and not only entanglement. In mathematical terms the quantum discord ${D}_{13}(\rho_{13})$ is defined as the difference of the quantum mutual information $I(\rho_{13})$, which contains all the classical and quantum correlations between the subsystems, and the classical correlations $J(\rho_{13})$, which contains all the information that can be obtained through local measurements on one of the subsystems. 
\begin{eqnarray}
&&{D}_{13}(\rho_{13})=I(\rho_{13})-\text{max}_{\hat{\Pi}_1}\left\{J(\rho_{13})\right\}
\\ &&
I(\rho_{13})=S(\rho_{1})+S(\rho_{3})-S(\rho_{13})
\\ &&
J(\rho_{13})=S(\rho_1)-S(\rho_{13 \vert \hat{\Pi}_1^\alpha})
\label{classical}
\end{eqnarray}

Where $S(\rho)=-\trace{\rho \log\left(\rho\right)}$ is the Von Neumann Entropy. The reduced density matrices $\rho_1, \rho_3$ are obtained after taking the partial trace of the state $\rho_{13}$ with respect to subsytem 3 and 1 respectively. 
\begin{eqnarray}
\rho_1=\rho_3=\frac{1}{2}\begin{pmatrix}
		 1+r & 0 \\
		 0 & 1-r \\
		  \end{pmatrix}
\end{eqnarray}

With the following definition, and the  eigenvalues $\Lambda_i$ of $\rho_{13}$ the mutual information can be written as:
\begin{eqnarray}
&& h(x)=\frac{1+x}{2}\log{\left(\frac{2}{1+x}\right)}+\frac{1-x}{2}\log{\left(\frac{2}{1-x}\right)} \ \ \ 
\\ &&
I(\rho_{13})=2h(r)+\sum_{i=1}^4 \Lambda_i \log{\left(\Lambda_i\right)}
\label{mutual}
\end{eqnarray}

To calculate the Discord ${D}_{13}(\rho_{13})$ it is necessary a maximization over all possible measurements $\hat{\Pi}_1$ over subsytem 1; the maximization is introduced to eliminate any dependence of the Discord with respect to the measurements.  The calculation of the classical correlations $\text{max}_{\hat{\Pi}_1}\left\{J(\rho_{13})\right\}$ is similar to the one on reference \cite{Frey}. Let it be $\hat{\Pi}_1^\alpha$ the local projection measurements to be applied on qubit 1. 
\begin{eqnarray*}
&&
\hat{\Pi}_1^\alpha=\frac{1}{2} \left(\mathcal{I}_1+\alpha \ \uv{r}_1 \cdot \gv{\sigma_1}\right) \ \ \ \alpha=\pm1
\\ &&
\uv{r}_1=\left(\sin{(\theta)}\cos{(\phi)},\ \sin{(\theta)}\sin{(\phi)}, \ \cos{(\theta)}\right) 
\end{eqnarray*}

The maximization over the measurements $\hat{\Pi}_1$ implies by equation (\ref{classical}) the calculation of the minimum: 
 \begin{eqnarray*}
 && \text{min}_{\hat{\Pi}_1}\left\{ S(\rho_{13 \vert \hat{\Pi}_1^\alpha})\right\}=\text{min}_{\hat{\Pi}_1^\alpha}\sum_\alpha q_\alpha S(\rho_\alpha)
\\ &&
\rho_\alpha=\frac{1}{q_\alpha}\left(\hat{\Pi}_1^\alpha\otimes\mathcal{I}_3\right)\rho_{13}\left( \hat{\Pi}_1^\alpha\otimes\mathcal{I}_3\right)
\\ &&
q_\alpha=\frac{1}{2} \left(1+\alpha r \cos(\theta)\right)
\end{eqnarray*}

The eigenvalues $\Lambda_{\alpha}$of $\rho_\alpha$ were found to be:
\begin{widetext}
\begin{eqnarray*}
\Lambda_\alpha=\left\{0, 0, \frac{1}{2}\pm \frac{\sqrt{f(\theta, \phi)+\alpha \ 2 r \ c_3 \ \cos{(\theta)}}}{4 q_\alpha}\right\}
\ \ \ \ \text{where}  \ \ \ \ \ f(\theta, \phi)=r^2+c_3^2 \cos{(\theta)}^2 +\sin{(\theta)}^2\left(c_1^2 \cos{(\phi)}^2-c_2^2 \sin{(\phi)}^2\right)
\end{eqnarray*}
\end{widetext}

The minimization was done through the calculation of the Hessian Matrix and numerical calculations. The minimization is achieved when $\phi=0$ and $\theta=\pi/2$ for all values of $\kappa , \lambda $ and $T$. Therefore the classical correlations $\text{max}_{\hat{\Pi}_1}\left\{J(\rho_{13})\right\}$ can be written as: 
\begin{eqnarray}
 && \text{min}_{\hat{\Pi}_1}\left\{ S(\rho_{13 \vert \hat{\Pi}_1^\alpha})\right\}= h(\sqrt{r^2+c_1^2})
\\ &&
\text{max}_{\hat{\Pi}_1}\left\{J(\rho_{13})\right\}=h(r)-h(\sqrt{r^2+c_1^2})
\label{classical2}
\end{eqnarray}

The Quantum Discord ${D}_{13}(\rho_{13})$ will be obtained by the sustraction of the mutual information (\ref{mutual}) and the classical correlations (\ref{classical2}). 
\begin{equation}
{D}_{13}(\rho_{13}) =h(r)+\sum_{i=1}^4 \Lambda_i \log{\left(\Lambda_i\right)}+h(\sqrt{r^2+c_1^2})
\end{equation}
A similar calculation can be done to obtain the quantum correlations when the measurements are done on qubit 3. It was found that the Discord is symmetric for the system defined by $\rho_{13}$. 
\begin{equation}
{D}_{13}(\rho_{13}) ={D}_{31}(\rho_{13})
\end{equation}
    \acknowledgments
	We want to express our thanks to Michael R. Frey from Bucknell University for his comments regarding the paper's structure. Jose Trevison acknowledges that this research has been partially supported by the Ministry of Education, Culture, Sports, Science and Technology of Japan  (MEXT) under the scholarship No. 133058.

	

		\end{document}